\documentclass[aps,pre,epsfigure,superscriptaddress,nofootinbib,notitlepage]{revtex4-1}
\usepackage[colorlinks=true,linkcolor=blue,urlcolor=blue,citecolor=blue,pdfusetitle]{hyperref}
\usepackage[utf8]{inputenc}
\usepackage[english]{babel}
\usepackage{amsmath}
\usepackage[caption = false]{subfig}
\usepackage{verbatim}
\usepackage{graphicx,epstopdf}
\usepackage{blindtext}
\usepackage[table,xcdraw]{xcolor}
\usepackage{lipsum}
\usepackage{amsfonts}
\usepackage{bbm}
\usepackage{amssymb}
\usepackage{enumerate}
\usepackage{color}
\usepackage{latexsym}
\usepackage{physics}
\usepackage{times,txfonts}
\usepackage{braket}
\usepackage{soul}

\newcommand{\dbar}{d\hspace*{-0.08em}\bar{}\hspace*{0.1em}}

\newcommand{\Dcal}{\mathcal{D}}
\newcommand{\Ecal}{\mathcal{E}}

\newcommand{\Hcal}{\mathcal{H}}

\newcommand{\1}{\mathbbm{1}}

\usepackage{setspace} 

\begin{document}

\doublespacing  

\title{Quantum Thermodynamics in Spin Systems: A Review of Cycles and Applications}

\author{Vinicius Gomes de Paula}
\email{viniciusgomespaula@id.uff.br}
\affiliation{Institute of Physics, Fluminense Federal University, Av. Gal. Milton Tavares de Souza s/n, 24210-346 Niterói, Rio de Janeiro, Brazil.}

\author{Wanisson S. Santana}
\email{wanisson.santana@ufob.edu.br}
\affiliation{Institute of Physics, Fluminense Federal University, Av. Gal. Milton Tavares de Souza s/n, 24210-346 Niterói, Rio de Janeiro, Brazil.}
\affiliation{Quantum Information and Statistical Physics Group, Western Bahia Federal University - Campus Reitor Edgard Santos. Bertioga Street, 892, Morada Nobre I, 47810-059 Barreiras, Bahia, Brazil.}

\author{Clebson Cruz}
\email{clebson.cruz@ufob.edu.br}
\affiliation{Quantum Information and Statistical Physics Group, Western Bahia Federal University - Campus Reitor Edgard Santos.  Bertioga Street, 892, Morada Nobre I, 47810-059 Barreiras, Bahia, Brazil.}

\author{Mario Reis}
\email{marioreis@id.uff.br}
\thanks{Permanent address: Fluminense Federal University, Brazil; Temporary address (hosting during the writing of the present work): University of Aveiro, Portugal.}
\affiliation{Department of Physics and CICECO, University of Aveiro, Universitary Campus of Santiago, 3810-193 Aveiro, Portugal}
\affiliation{Institute of Physics, Fluminense Federal University, Av. Gal. Milton Tavares de Souza s/n, 24210-346 Niterói, Rio de Janeiro, Brazil.}

\begin{abstract}
Quantum thermodynamics is a powerful theoretical tool for assessing the suitability of quantum materials as platforms for novel technologies. In particular, the modeling of quantum cycles allows us to investigate the heat changes and work extraction at the nanoscale, where quantum effects dominate over classical ones. In this Review, we cover the mathematical formulation used to model the quantum thermodynamic behavior of small-scale systems, building up the quantum analog versions of thermodynamic processes and reversible cycles. We discuss theoretical results obtained after applying this approach to model Heisenberg-like spin systems, which are toy models for metal complex systems. In addition, we discuss recent experimental advances in this class of materials that have been achieved using the quantum thermodynamic approach, paving the way for the development of quantum devices. Finally, we point out perspectives to guide future efforts in using quantum thermodynamics as a reliable and effective approach to establish metal complex systems as quantum materials for energy storage/harvesting purposes.
\end{abstract}


\maketitle

\tableofcontents

\section{Introduction}\label{sectionI}

Quantum energy extraction approaches have been developed in recent years due to the emergence of the so-called quantum energy science \cite{metzler2023emergence}. In this context, thermodynamics plays a crucial role due to its robust predictions for classical and quantum systems \cite{campaioli2023colloquium}. Initially, incorporating quantum effects appeared to have a negligible impact on traditional thermodynamic principles \cite{PRL2013Huber,campaioli2023colloquium}. However, equilibrium thermodynamics does not impose restrictions on the rate of energy transformations, such as the conversion of energy into heat and work, which has sparked significant interest in the use of quantum systems as thermodynamic devices \cite{myers2021quantum}. This interest has led to the emergence of the field of \textit{Quantum Thermodynamics}, which investigates the dynamics of heat and work at the quantum scale, offering a rigorous mathematical framework to model thermodynamic processes and reversible cycles within quantum systems \cite{Gemmer:Book}. In this context, the understanding of quantum cycles can also shed light on the development of more efficient energy conversion technologies and advanced quantum systems \cite{campaioli2023colloquium,metzler2023emergence,auffeves2022quantum,myers2022quantum}, allowing to investigate work extraction of small-scale quantum thermal machines based on real materials \cite{cruz2022quantum,cruz2023quantum}.

Due to the broadness and relevance of this topic, we present a review covering the mathematical framework required for modeling quantum energy extraction as well as its applications in the context of quantum magnetic materials. In section \ref{sec:QuanThermProc} we present mathematical expressions for quantum isothermal, isochoric, and adiabatic processes and build up different quantum cycles using as starting point the quantum analogs of heat and work  \cite{Alicki:79}. We also performed a thoroughly physical discussion of the cycles and the heat fluxes on each quantum process described here. 
The section \ref{sec:LitRes} is devoted to illustrating the application of this theory. We discuss collected literature reports for which interacting spin configurations have been used as working substances for modeling Carnot, Stirling, and Otto small-scale thermal machines. Since these Heisenberg-like spin configurations are toy models for metal complex systems, we have also added a handful of experimental literature results obtained on real metal complex systems, ranging from experimental implementation of quantum cycles up to the development of quantum battery devices. Characterizing the quantum correlations in metal complex systems is crucial to understanding how these resources can be effectively used in applications, since their robust quantum properties, such as long coherence times \cite{kazmierczak2021impact, jiang2021exotic, kazmierczak2021impact, moreno2021measuring,moreno2018molecular,coronado2020molecular} and room temperature quantum correlations \cite{cruz2016carboxylate,cruz2022quantum}, are highly relevant for the development of quantum devices \cite{coronado2020molecular, moreno2023single, zhang2024molecular, cruz2017influence,kova2020unconventional,wasielewski2020exploiting,gaita2019molecular,mezenov2019metal,moreno2018molecular,moreno2021measuring,chakraborty2014signature,szalowski2022two,sahling2015experimental,chakraborty2013experimental,PhysRevB.103.224434,PhysRevLett.127.037201,PhysRevResearch.2.043329}. From numerical modeling of Heisenberg-like spin configurations up recent experimental advances upon experimental validation of these cycles, we discuss the impact of such findings to the establishment of novel quantum energy technologies. Next, in section \ref{sec:Persp}, we comment on some perspectives for the future exploration of metal complex systems as platforms for the development of emerging technologies on the quantum energy extraction scenario, paving the way for the development of room temperature quantum energy devices \cite{hoffmann2022quantum, tanaka2023metal} Finally, in section \ref{sec:append} we describe in detail the mathematical models used to obtain all the expressions that quantify the heat changes during the quantum thermodynamic processes and cycles, showing that they all obey the First Law of thermodynamics.

\section{Quantum thermodynamics processes and cycles}\label{sec:QuanThermProc}

Fast miniaturization of devices that exploit quantum properties \cite{millen2016rise, merali2017bending, kumar2023introduction} has led to the development of the quantum thermodynamics field, which main objective is to investigate the thermodynamic behavior of small-scale systems. In the next sections, we will present the main concepts of the theory, defining quantum analogs for the First Law of thermodynamics, thermodynamic processes, and reversible thermodynamic cycles. Also, we obtain mathematical expressions for all the relevant thermodynamic quantities, such the heat changes in each process, the net Work per cycle, and the efficiency of the system, in terms of the partition function of thermalized states.

\subsection{First-Law of quantum thermodynamics}

The Hamiltonian operator of a general quantum system is given by\cite{Deffner-Campbel:Book}:
\begin{equation} \label{Hamiltonian_lambda}
    \hat{H}(\lambda)=\sum_n E_n (\lambda)\ket{n}\bra{n},
\end{equation}
where $\lambda$ is a parameter, $\{\ket{n}\}$ are the orthonormal eigenvectors associated with the eigenvalues $E_n (\lambda)$, and $\ket{n}\bra{n}$ are the outer products forming projection operators. At this point, it is important to mention that there is a quantum analog of the quasi-static condition in classical thermodynamics, which is labeled as ultra-weak coupling \cite{Deffner-Campbel:Book}. From a physical point of view, an ultra-weak coupling between the system and its thermal environment means that, upon changes in the bath temperature, the system quickly reaches thermal equilibrium. This is because the time scale of the
variation of the work parameter is considered to be much shorter than the relaxation time. 
In particular, in the context of metal complex systems, the quasi-static process can be carried out in an experimental setup by making an infinite time approximation \cite{cruz2022}. As the quasi-static model does not consider time, the static magnetic measurements can be utilized to demonstrate the compatibility of the offered theoretical models with an actual situation. Each data point is acquired within seconds, but the spin-spin relaxation times for these metal complexes, as mentioned in references \cite{moreno2018molecular,gaita2019molecular,moreno2021measuring,coronado2020molecular}, typically range from $1$ to $100$ microseconds. Therefore, the thermal baths can be utilized by taking into account the temperature of the measuring devices, as the timescale of the measurement is usually much longer than the time it takes for the system to reach equilibrium \cite{cruz2022}. An experimental case study is discussed in details in section \ref{sec:QuanThermCyc}.

Such equilibrium states are described by the density operator $\hat{\rho}_{eq}(\lambda)$, which is written as:
\begin{equation} \label{pho_lambda}
    \hat{\rho}_{eq}(\lambda)=\sum_{n}p_n(\lambda)\ket{n}\bra{n},
\end{equation}
where $p_n(\lambda)=e^{-E_n(\lambda)/k_BT}/Z$ is the occupation probability of the $n$-th energy state and $Z$ is the partition function. Equation \ref{pho_lambda} describes an equilibrium state (Gibbs state) analog to the classical ones defined under the thermodynamic limit ($N$ particles occupying a volume $V$ and making $N \rightarrow \infty$ and $V \rightarrow \infty$, while $N/V$ reamins finite). 

The internal energy of the system is the quantum-statistical average over the Hamiltonian:
\begin{equation} \label{energy_U}
 U =Tr\{\hat{\rho}_{eq}\hat{H}\}=   \sum_{n} p_nE_n.
 \end{equation}
Differentiating equation \ref{energy_U}, we obtain:
\begin{equation} \label{diff_energy_U}
  dU = \sum_{n} p_n dE_n + \sum_{n} E_n dp_n,
\end{equation}
Which can be interpreted as the First Law of quantum thermodynamics. This can be compared with its classical counterpart:
\begin{equation} \label{diff_First_Law}
  dU = \dbar W + \dbar Q  = \sum \mathcal{F}_id\mathcal{X}_i + TdS,
\end{equation}
where $S$ refers to the entropy, $\mathcal{X}_i$ is the generalized coordinate and $\mathcal{F}_i$ is the corresponding force. For instance, considering a confined gas by adiabatic walls, with a movable piston, the generalized force is the pressure $P$ upon the piston, and the generalized coordinate is its position $x$. According to the classical definition, the work done on or by the system is defined as the change of internal energy produced by the modifications on the generalized coordinates. In quantum mechanics, these variations are inherently linked to alterations in the energy levels of the system \cite{Deffner-Campbel:Book}. According to this definition, heat will be the result of the change in the internal energy originating from the alterations in the probabilities of occupation of the energy levels. Furthermore, this interpretation is not an extension of the classical concept of heat, which refers to the alteration in internal energy followed by a change in entropy. However, it is a valid interpretation, particularly for systems in internal equilibrium where $\left[ \Hcal, \rho\right] = 0$ \cite{Deffner-Campbel:Book}. 

Thus, $dW$ is related to changes in the energy levels $E_n$ of the system and allows us to identify the first term of the equation \ref{diff_energy_U} as $dW$ the quantum work done:
\begin{equation} \label{quantum_dW}
  dW = \sum_{n} p_n dE_n.
\end{equation}
Where $dW>0$ ($dW<0$) stands for work performed on (by) the system. Such signal convention notation stands from the point of view of the working substance, as will be highlighted in section \ref{sec:QuanThermCyc}. Note that the quantum definition of work follows the classic thermodynamics (equation \ref{diff_First_Law}), where $dW$ can only be performed through changes in the generalized coordinates of the system, that change the energy levels $E_n$ of quantum systems \cite{Kieu:04, Schrodinger:Book}. 

On the other hand, changes in the populations $p_n$, change the Shannon entropy:
\begin{equation} \label{Shannon_entropy}
  S  = k_B \sum_n p_n\ln p_n.
\end{equation}
For the classical case, the entropy change is related to the heat exchange through the relation $\dbar Q=TdS$. Then, for the quantum case, we conclude that $dQ$ is associated with changes in the occupancy probability of each energy level $p_n$. It allows us to identify the second term of the equation \ref{diff_energy_U} as $dQ$, the quantum heat exchange:
\begin{equation} \label{quantum_dQ}
  dQ = \sum_{n} E_n dp_n,
\end{equation}
where $dQ>0$ ($dQ<0$) means heat absorbed (released) by the system, adopting the point of view from the working substance.

\subsection{Quantum Thermodynamic Processes}

\begin{figure*}[ht]
		\centering
		\includegraphics[width=\textwidth]{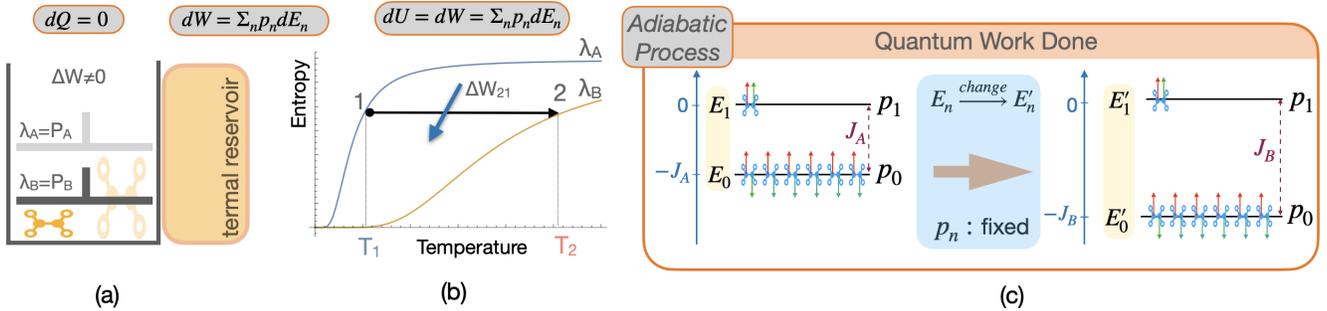}
		\caption{ Schematic representation of the quantum adiabatic process for a hypothetical dimer that constitutes a two-level system. (a) For this process, there is only work hence the occupation probabilities $p_0$ e $p_1$ are kept fixed, while the energy values $E_0$ and $E_1$ change to $E_0'$ and $E_1'$, respectively. (b) In an idealization of the quantum adiabatic process, the hypothetical dimer inside the container has no thermal contact with the thermal reservoir, ensuring no heat transfer. The external parameter (pressure) changes between $P_A$ and $P_B$ values, changing the magnetic coupling constant values between $J_A$ and $J_B$. This change in the pressure is responsible for the work done in the process. The sign of work is determined by whether the variation of the working parameter is performed by the system or on the system. As we can see, under compression, it results in work done on the system. (c) The Entropy $S$ as a function of the Temperature $T$ shows the corresponding path followed by the quantum adiabatic process. Since the occupation probabilities remain fixed, the system's temperature changes between $T_1$ and $T_2$, but its entropy remains constant. Thus, as in the classical counterpart, the quantum adiabatic is also an isentropic process.}
    \label{quantum_adiabatic}
\end{figure*}

\begin{figure*}[ht]
		\centering
		\includegraphics[width=\textwidth]{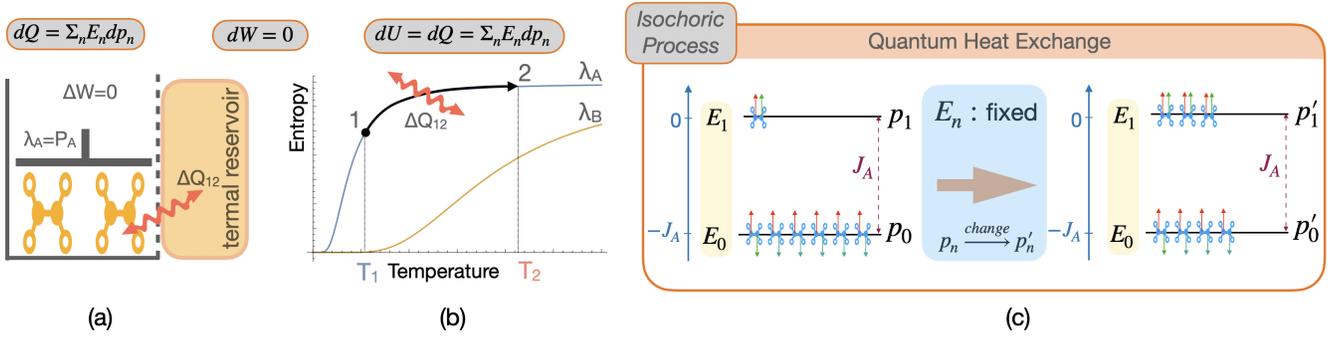}
		\caption{Schematic representation of the quantum isochoric process for a hypothetical dimer that constitutes a two-level system. (a) For this process there is only heat change, hence the occupation probabilities $p_0$ e $p_1$ change into $p_0'$ and $p_1'$ respectively, while the energy values $E_n$ are kept fixed. (b) In an experimental idealization of the quantum isochoric process, the hypothetical dimer is put inside a container in thermal contact with a thermal reservoir, which allows a heat exchange at constant pressure $P_A$ that ensures the magnetic coupling constant is fixed at a value $J_A$. (C) The Entropy $S$ as a function of the Temperature $T$ shows the corresponding path for this process. The Entropy changes since the occupation probabilities are not constant, but the path remains at a constant value $\lambda_A$ of the external parameter.  ) }
    \label{quantum_isochoric}
\end{figure*}

\begin{figure*}[ht]
		\centering
		\includegraphics[width=\textwidth]{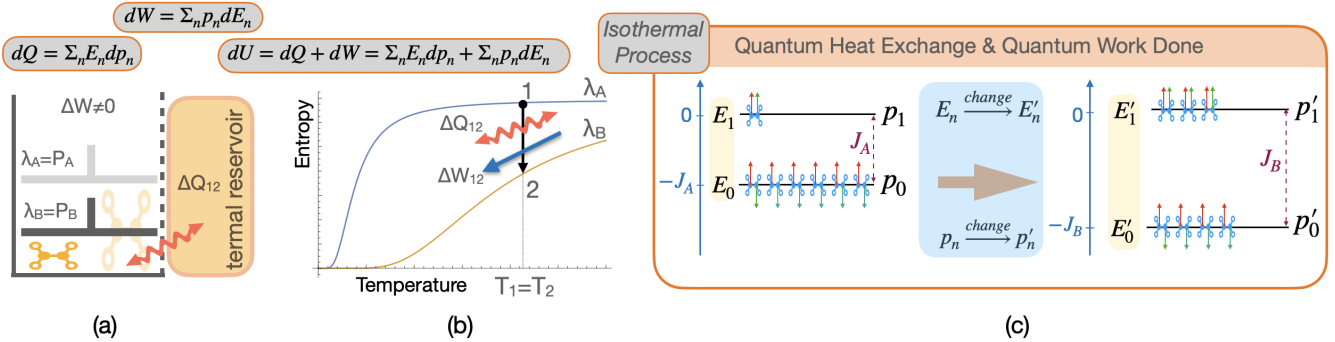}
		\caption{Schematic representation of the quantum isothermal process for a hypothetical dimer that constitutes a two-level system. (a) For this process, the temperature is kept fixed while the energy levels change from $E_0$ and $E_1$ and the occupation probabilities change from $p_0$ and $p_1$ to $p_0'$ and $p_1'$. The variation in energy levels and occupation probabilities leads to changes in the system's internal energy, in contrast to its classical counterpart. (b) In an experimental idealization of the quantum isothermal process, the hypothetical dimer inside the container is in contact with the thermal reservoir, ensuring heat transfer. Simultaneously, the pressure changes between $P_A$ and $P_B$ values changing the magnetic coupling constant values between $J_A$ and $J_B$, and there is work done. This change in the pressure is responsible for the work done in the process. The sign of work is determined by whether the variation of the working parameter is performed by the system or on the system. As we can see, the system is being compressed, resulting in work being done on the system. (c) The Entropy $S$ as a function of the Temperature $T$ shows the corresponding path followed by the quantum isothermal process. The Entropy changes while the external parameter also changes its values. The direction of the heat flux is determined by whether the system absorbs or rejects heat during the process.
  }
    \label{quantum_isothermal}
\end{figure*}

Thermodynamics processes are performed under a physical constraint (fixed temperature or entropy, for example), resulting in energy exchanges. 
By combining the processes adequately, it is possible to build up thermodynamic cycles to extract energy from quantum physical systems. In this section, we describe the quantum version of these thermodynamic steps, considering a dimer metal complex for the spins interact via a magnetic coupling constant $J$ that is sensitive to an external pressure $P$. For this two-level quantum system, the  energy levels are labeled as $E_0$ and $E_1$ while the corresponding probability occupations are denoted by $p_0$ and $p_1$.

\subsubsection{Quantum adiabatic process} \label{sec:qap}

This process is characterized by zero heat exchange between the working substance and the thermal reservoir, and therefore $dQ=0$. From the First Law of quantum thermodynamics (equation  \ref{diff_energy_U}), it is possible to write:
\begin{equation}\label{adiabatic_eq}
    dU=dW=\sum_np_ndE_n,
\end{equation}
which can conveniently be written as a function of the partition function. Considering a dinuclear metal complex model under mechanical constraint, the material shall be under pressure, to guarantee the work done the changes in the $\lambda$ external parameter. In a context where $\lambda$ stands for mechanical pressure (see Figure \ref{quantum_adiabatic}(a)), such variation changes the magnetic coupling $J$ of the system \cite{cruz2022}. In this case, there is no heat exchange and a certain amount of work is done. Let's now focus our attention on the entropy \textit{vs.} temperature diagram, as shown in Figure \ref{quantum_adiabatic}(b): the process goes from point (1) to point (2) (or vice-versa), where these two points belong to the entropy curves of pressure $P_A$ and $P_B$, changing the temperature from $T_1$ to $T_2$ (or vice-versa). Considering the step goes from $1\rightarrow 2$, for instance, the working substance starts at $T_1$, with energy spectra $E_{n}$. Work is done and the system reaches thermal equilibrium at temperature $T_2$, with energy spectra $E_{n}^\prime$. During the thermal equilibrium processes, there is no heat exchange, considering the occupation probabilities remain unchanged. The partition function for these two points are: $Z_1(T_1,E_{n})$ and $Z_2(T_2,E_{n}^\prime)$. Therefore equations \ref{adiabatic_eq} and \ref{energy_U} allow us to write \cite{chatterjee2021temperature, thomas2019quantum}: 
\begin{align} \label{W_21_adiabatic}
   \nonumber \Delta W_{21} &=\Delta U_{21}=U_{2} - U_{1}\\ &= k_B T_2^2 \frac{\partial }{\partial T_2}\ln\left[Z_{2}(T_2,E_{n}^\prime)\right]- k_B T_1^2 \frac{\partial }{\partial T_1}\ln \left[Z_{1}(T_1,E_{n})\right].
\end{align}

Finally, this process can also be seen in Figure \ref{quantum_adiabatic}(c), where the energy levels and their corresponding level occupancy of the hypothetical dimer are displayed. Considering there is no heat exchange and only work is done, the energy levels change due to external excitation, keeping unchanged the occupation probability for the levels. For the dimer of our example, the exchange interaction changes from $J_A$ to $J_B$,  due to the external pressure. It is important to highlight that, in classical thermodynamics, the adiabatic condition is achieved for fast processes, such as a free expansion or compression of a gas. For its quantum counterpart, the process must be slow enough to keep the occupancy probabilities $p_n$ unchanged \cite{Quan:07} as the system runs from the initial to the final states. Equation \ref{W_21_adiabatic} writes the work done considering the partition function of the system, and this partition function depends on the energy spectra change (due to work done) and temperature (to keep the occupancy probability constant). The partition function then can be determined case-by-case.

\subsubsection{Quantum isochoric process} \label{sec:isochoric}

The classical isochoric process is characterized by constant volume $V$ of the system, while temperature $T$ and pressure $P$ change, resulting in $dW=0$. From the First Law of quantum thermodynamics (equation  \ref{diff_energy_U}), it is possible to write:
\begin{equation}\label{isochoric_eq}
    dU=dQ=\sum_nE_ndp_n,
\end{equation}
which can also conveniently be written as a function of the partition function of the system.
For our hypothetical dimer under mechanical constraint, the material would be at constant pressure to guarantee the condition of no work done. See Figure \ref{quantum_isochoric}(a), where the metal complex (dimer) can exchange heat $\Delta Q_{12}$ with the thermal reservoir. We can see this process from another perspective: an entropy \textit{vs.} temperature diagram, as shown in Figure \ref{quantum_isochoric}(b). Note the process follows the $\lambda_A=P_A$ (constant pressure) path, exchanging $\Delta Q_{12}$ amount of heat with the reservoir and, consequently, changing the temperature from $T_1$ to $T_2$ (or vice-versa). The step $1\rightarrow 2$, for instance, the working substance starts at $T_1$, with energy spectra $E_{n}$, and exchanges heat with the reservoir until it reaches thermal equilibrium at temperature $T_2$, with the same energy spectra $E_{n}$. The partition functions for these two points are $Z_1(T_1)$ and $Z_2(T_2)$, and these have no explicit dependence on the energy spectra of the system. Considering no work is done, the heat exchange equals the change in the internal energy. Therefore equations \ref{energy_U} and \ref{isochoric_eq} allow us to write \cite{chatterjee2021temperature, thomas2019quantum}: 
\begin{align} \label{Q_21_isochoric}
    \nonumber \Delta Q_{21} &= \Delta U_{21}=U_{2} - U_{1} \\
    &= k_B T_2^2 \frac{\partial }{\partial T_2}\ln\left[Z_{2}(T_2)\right]- k_B T_1^2 \frac{\partial }{\partial T_1}\ln\left[Z_{1}(T_1)\right]
\end{align}
for an isochoric quantum process.

We can also see this process in Figure \ref{quantum_isochoric}(c), presenting the energy levels of the hypothetical dimer. There is no work done for this isochoric process and only heat exchange, depending on the level's occupation change (see equation \ref{isochoric_eq}). For this example, we are dealing with the exchange interaction $J_A$ that does not change along the process (constant pressure) and the singlet-triplet gap that remains unchanged. On the other hand, the populations for these two levels change accordingly. It is important to mention that other physical stimuli than pressure could be considered, such as an external magnetic field.

\subsubsection{Quantum isothermal process}\label{sec:isothermal}

For this process, such as in its classical counterpart, the system maintains a constant temperature $T$, however, the energy levels $E_n$ and the occupation probabilities $p_n$ change simultaneously \cite{Quan:07}. From the First Law of quantum thermodynamics (equation  \ref{diff_energy_U}), it is possible to write:
\begin{equation}\label{adiabatic_eq2}
    dU=dQ+dW=\sum_nE_ndp_n+\sum_np_ndE_n,
\end{equation}
which can be written as a function of the partition function, considering the Shannon entropy written as:
\begin{equation} \label{first_law_modified}
    ST = U - k_B T\ln Z.
\end{equation}
From the dimer under mechanical constraint we are discussing, the material shall be under pressure, to guarantee the work done. This process also exchanges heat $\Delta Q_{12}$ with the thermal reservoir. See Figure \ref{quantum_isothermal}(a). Let's now see the entropy \textit{vs.} temperature diagram, as shown in Figure \ref{quantum_isothermal}(b): the process goes from point (1) to point (2) (or vice-versa), where these two points belong to the entropy curves of pressure $P_A$ and $P_B$, keeping a constant temperature $T_1=T_2$. For the step $1\rightarrow 2$, for instance, the working substance starts from point 1 with energy spectra $E_{n}$. The final energy spectra is $E_{n}^\prime$. During the isothermal process, there is heat exchange due to the changes in the occupation probabilities. The partition function for these two points are: $Z_1(E_{n})$ and $Z_2(E_{n}^\prime)$, and these have no explicit temperature dependence, considering $T_1=T_2$. Considering the entropy given by equation \ref{first_law_modified}, it is possible to calculate the heat exchange during the two isothermal processes as \cite{chatterjee2021temperature, thomas2019quantum}:
\begin{equation} \label{Q_21_isothermal}
        \Delta Q_{21} = Q_2-Q_1 = T_2(S_2-S_1)=(U_2-U_1)-k_BT_2\ln\left[\frac{Z_2(E_{n}^\prime)}{Z_1(E_{n})}\right],
\end{equation}
where the first term of the above equation represents the internal energy change, while the second term is the work done. The above equation can be rewritten as:
\begin{equation}\label{Q_21_isothermal_2}
  \Delta Q_{21} =\underbrace{k_B T_2^2 \frac{\partial}{\partial T_2} \ln\left[ \frac{Z_{2}(E_{n}^\prime)}{Z_{1}(E_{n})}\right]}_{\Delta U_{21}} - \underbrace{k_B T_2 \ln \left[ \frac{Z_2(E_{n}^\prime)}{Z_1(E_{n})} \right]}_{\Delta W_{21}}.
\end{equation}

Finally, this process can also be seen in Figure \ref{quantum_isothermal}(c), where the energy levels of the dimer are shown, as well as the corresponding hypotheticalal level occupancy. Considering there is heat exchange and work done simultaneously, the occupation probability and the energy levels change due to external excitation. For the dimer of our example, due to the external pressure, the exchange interaction changes from $J_A$ to $J_B$ (or vice versa), also changing the occupancy probability of the energy levels - in such a way that the system's temperature remains unchanged. 

\subsection{Quantum Thermodynamic Cycles}\label{sec:QuanThermCyc}

This section discusses the quantum processes previously described to build up a few quantum thermodynamic cycles. The approach here is suitable to predict the thermodynamic properties of real metal complex materials.  With this in mind, after presenting each cycle we discuss literature results describing quantum thermodynamic cycles of distinct spin system for which the Hamiltonians used are toy models for the metal complexes (dimer, trimer, etc) 

\subsubsection{Efficiency, Coefficient of performance and Operation Modes}\label{sec:ecop}

\begin{figure}[ht]
		\centering
	\includegraphics[width=0.3\textwidth]{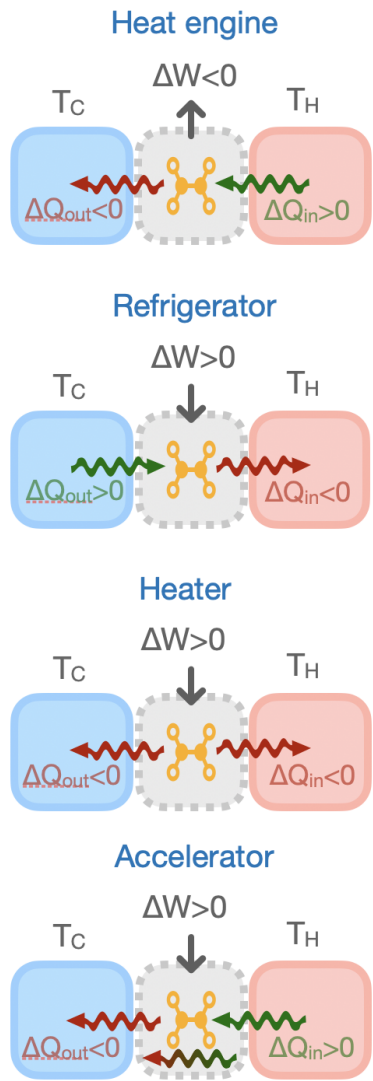}
		\caption{Schematic representation of the operation modes allowed by the 2º Law of thermodynamics. The heat engine is characterized by the system absorbing heat $Q{in} > 0$ from the hot bath at $T_{H}$ and releases heat $Q_{out} > 0$ into the cold bath at $T_{C}$; the difference between both heat fluxes is converted into work done by the system $(\Delta W < 0)$. In the refrigeration mode, both heat fluxes are inverted, therefore the system releases heat  $\Delta Q{in} > 0$ at the hot bath $T_{H}$ while absorbing heat $\Delta Q{out} > 0$ from the cold bath $T_{C}$; and in this case the difference between both heat fluxes gives rise to a work performed on the system $(\Delta W > 0)$. The heater mode is characterized by the system releasing heat to both thermal baths ($\Delta Q{in} < 0$ and $\Delta Q{out} < 0$) resulting in work performed on the system ($\Delta W > 0$), i.e. the system uses the work to induce heat flux from in both thermal baths. Finally, the accelerator mode exploits the work done on the system ($\Delta W > 0$) to enhance the heat flux from the hot bath $T_{H}$ to the cold bath $T_{C}$. }
    \label{fig:modes}
\end{figure}

According to the Second Law of thermodynamics \cite{solfanelli2020nonadiabatic, myers2021quantum}, thermodynamic cycles can display four distinct operation modes: heat, refrigerator, accelerator, and heater. The classification of each operation mode depends on the signs of the heat exchanged between the working substance and the hot reservoir  $\Delta Q_{in}$, and the cold reservoir $\Delta Q_{out}$, as well as the net work done $\Delta W$ on or by the working substance. At this point, it is worth highlighting the signal convention of the energy exchange. From the viewpoint of the working substance, when it absorbs heat, the system experiences an increase in energy, which is supposed to be positive. Conversely, when the system releases heat, it loses energy, which is expected to be negative. In the same way, when the system performs work, it is expected to have a negative value due to the loss of energy. Conversely, if work is done on the system, it is expected to have a positive value. Table \ref{tab:modes} describes physically these operation modes and the corresponding signs for the $\Delta Q_{in}$, $\Delta Q_{out}$, and $\Delta W$ quantities that define each one of them. Complementarily, Figure \ref{fig:modes} displays a visual representation of each of the operation modes, highlighting the directions of the heat flux on each case, considering the thermal baths $T_{C}$ and $T_{H}$.
\begin{table*}[ht]
\caption{Relationship of the distinct operation modes allowed by the 2º Law of thermodynamics and the signs of the relevant quantities $\Delta W$, $\Delta Q_{in}$ and $\Delta Q_{out}$, following the signal convention considering the point of view from the working substance. \label{tab:modes}}
\begin{tabular}{@{}c|c|c|c|c@{}}
\multicolumn{1}{c|}{\begin{tabular}[c]{@{}c@{}}Operation \\ mode\end{tabular}} & \multicolumn{1}{c|}{\begin{tabular}[c]{@{}c@{}}Net Work\\ $\Delta W$\end{tabular}} & $\Delta Q_{in}$ & $\Delta Q_{out}$ & Description \\ \hline\hline
Heat engine  &  - &  + &  - & Heat flows from the hot bath to the cold bath and work is done \textit{by} the system.\\
Refrigerator  &  + &  -  &  + & Heat flow is forced from the cold bath to the hot one due to work done \textit{on} the system.
\\
Accelerator &  +    &    +  &      -  & Heat flow is accelerated from the hot bath to the cold one due to work done \textit{on} the system.
\\
Heater   &   +   &         -        &      -  &  Heat flows in both baths, hot and cold, due to work done \textit{on} the system.
\end{tabular}
\end{table*}

While studying thermodynamic cycles we quantify the capacity of the working substance to transform energy into useful work through the efficiency $\eta$, which, for a heat engine, is defined as \cite{kumar2023introduction}:
\begin{equation}\label{eta_def}
    \eta = \frac{\Delta W}{\Delta Q_{in}}.
\end{equation}
On the other hand, if the workings substance operates in the refrigerator mode (see Table \ref{tab:modes} and/or Figure \ref{fig:modes}), the parameter to quantify its performance is the  coefficient of performance (COP), defined by \cite{el2023enhancing}:
\begin{equation}\label{cop_def}
    COP = \frac{\Delta Q_{in}}{\Delta W}.
\end{equation}

\subsubsection{Quantum Carnot cycle} 

\begin{figure}[ht]
		\centering
		\includegraphics[width=0.5\textwidth]{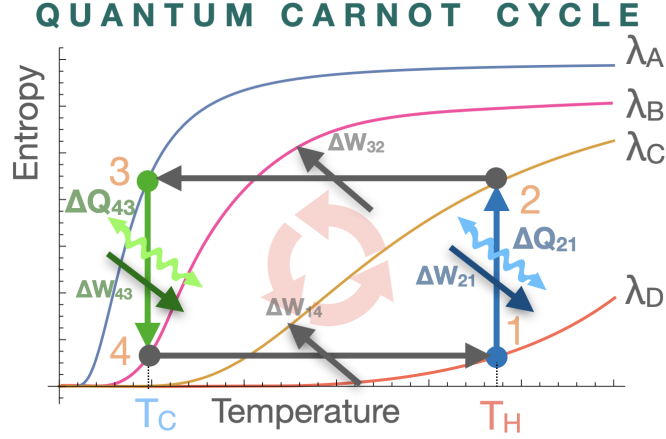}
		\caption{Entropy - Temperature S - T diagram of a quantum Carnot cycle. Note that the work is performed by the system ($W>0$) during the adiabatic cooling ($2 \rightarrow 3$) process, while it is performed on the system ($W<0$) during the adiabatic heating ($3 \rightarrow 4$) process. The heat exchange between the working substance and the reservoir occurs only during the adiabatic steps. Note that for each of the stages, the external parameter displays a specific distinct value. }
    \label{carnot}
\end{figure}

This cycle consists of two quantum isothermal and two quantum adiabatic processes. Figure \ref{carnot} shows each strike that composes the cycle, indicating the heat changes between the working substance and its surroundings. 

The amount of heat exchanged during the two isothermal strokes $1\rightarrow2$ and $3\rightarrow4$ are calculated via equation \ref{Q_21_isothermal}. On the other hand, as we have discussed along subsection \ref{sec:qap}, the strokes $2\rightarrow3$ and $4\rightarrow1$ corresponding to the quantum adiabatic processes, and do not produce heat. However, there is work done associated with these two strokes, calculated in accordance with equation \ref{W_21_adiabatic}. We note that a intrinsic characteristic of this cycle is that there is work during each of the four strokes.

Then, the net work during the cycle is computed as summing up the contributions from each stroke displayed in Figure \ref{carnot}:

\begin{align} \label{Wnet_Carnot}
\Delta W_{Carnot} 
&= \left(\Delta W_{21} - \Delta W_{43}\right)+ \left(\Delta W_{32} + \Delta W_{14}\right) \\
&= k_B T_H \ln \left[ \frac{Z_2(E_{n,2})}{Z_1(E_{n,1})} \right] -k_B T_C \ln \left[ \frac{Z_4(E_{n,4})}{Z_3(E_{n,3})} \right]\\
&+k_B T_C^2 \frac{\partial }{\partial T_C}\ln\left[Z_{3}(T_C,E_{n,3})\right]- k_B T_H^2 \frac{\partial }{\partial T_H}\ln \left[Z_{2}(T_H,E_{n,2})\right] \\
&+ k_B T_H^2 \frac{\partial }{\partial T_H}\ln \left[Z_{1}(T_H,E_{n,1})\right]-k_B T_C^2 \frac{\partial }{\partial T_C}\ln\left[Z_{4}(T_C,E_{n,4})\right]  \\ 
\end{align}

It is worth mentioning that to extract positive work from the bath using $\Delta W_{Carnot}$, we must impose the constraint $T_{H} > T_{C}$, as in the classical Carnot cycle \cite{Quan:07}.

Finally, the efficiency of the quantum Carnot cycle can be written as:
\begin{align}
    \label{eta_carnot}
 \nonumber  \eta_{Carnot} 
   &= 1 + \left( \frac{T_C}{T_H}\right)\frac{T_C \frac{\partial}{\partial T_C} \ln\left[ \frac{Z_{4}(T_C,E_{n,4})}{Z_{3}(T_C,E_{n,3})}\right] + \ln \left[ \frac{Z_4(E_{n,4})}{Z_3(E_{n,3})} \right]}{T_H \frac{\partial}{\partial T_H} \ln\left[ \frac{Z_{2}(E_{n,2})}{Z_{1}(E_{n,1})}\right] - \ln \left[ \frac{Z_2(E_{n,2})}{Z_1(E_{n,1})} \right]}   
\end{align}
If we impose the constraint given by:
\begin{equation} \label{constraint_carnot}
    \frac{Z_2(T_H,E_{n,2})}{Z_1(T_H,E_{n,1})}=\frac{Z_{4}(T_c,E_{n,4})}{Z_{3}(T_c,E_{n,3})}=c_{nte},
\end{equation}
we recover the classical Carnot limit:
\begin{equation} \label{eta_Carnot_clas}
    \eta_{Carnot}= 1 - \frac{T_C}{T_H}.
\end{equation}

Equation \ref{constraint_carnot} is equivalent to the adiabatic theorem, which ensures that adiabaticity (i.e. no changes in the $p_n$ values) of a quantum thermodynamic process. In other words, imposing the adiabatic theorem on the quantum Carnot cycle, we restrict it to operate within the classical Carnot limit as the upper bond.

\subsubsection{Quantum Stirling Cycle}

\begin{figure}[ht]
		\centering
	\includegraphics[width=0.5\textwidth]{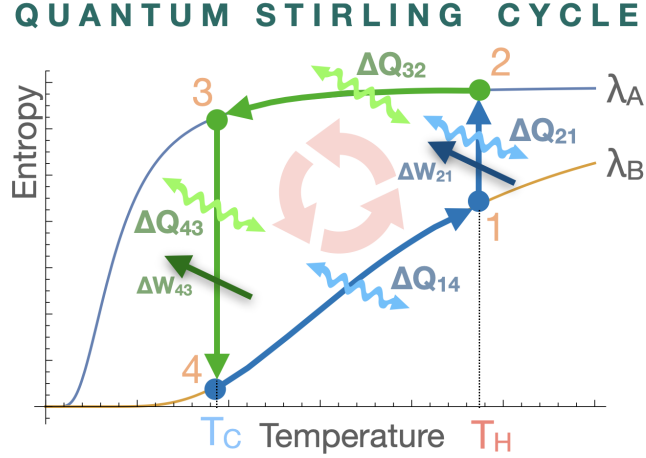}
		\caption{Entropy - Temperature S - T diagram of a quantum Stirling cycle. Note that the work is performed by the system ($W>0$) during the isothermal expansion ($1 \rightarrow 2$) process at equilibrium with the hot thermal bath $T_h$. In contrast, Work is performed on the system ($W<0$) during the isothermal compression ($3 \rightarrow 4$) process at thermal equilibrium with the cold bath $T_c$. Heat exchange between the working substance and the reservoir occurs during all four processes.}
    \label{stirling}
\end{figure}

By compromising two quantum isochoric and two isothermal processes it is possible to establish the quantum Stirling cycle. In this cycle, the working substance exchanges heat during all four strokes, as represented in Figure \ref{stirling}.

During the isothermal process $1 \rightarrow 2$ and $3 \rightarrow 4$, the working substance is coupled to a reservoir at temperature $T_H$ and $T_C$, respectively, while there is heat exchange and work done. Following what we have discussed along Section \ref{sec:isothermal}, it is possible to calculate the heat exchange during these two isothermal processes using equation \ref{Q_21_isothermal_2}. For the two isochoric processes $2 \rightarrow 3$ and $4 \rightarrow 1$, no work is done, as already discussed along Section \ref{sec:isochoric}, and the heat exchanged during these two processes can be obtained using equation \ref{Q_21_isochoric}. For details in these calculations refers to the appendix section \ref{AP:QuanStir}.

The net work of the quantum Stirling cycle is obtained summing the $\Delta W_{21}$ and $\Delta W_{43}$ quantities corresponding to the isothermal steps displayed in Figure \ref{stirling}:

\begin{align} \label{W_net_Stirling}
\Delta W_{Stirling} &= \Delta W_{21} -\Delta W_{43}\\
&= k_B T_H \ln \left[ \frac{Z_2(E_{n,2})}{Z_1(E_{n,1})} \right] -k_B T_C \ln \left[ \frac{Z_4(E_{n,4})}{Z_3(E_{n,3})} \right]
\end{align}
where we used the convention defined in Table \ref{tab:modes}: $\Delta W_{43}<0$ since it accounts for the Work performed \textit{on} the system while $\Delta W_{21}>0$ stand for work done \textit{by} the system.


From the total values of net work and heat exchanges, it is possible to calculate the efficiency of the Stirling cycle using equation \ref{eta_def}:
\begin{widetext}
{\begin{align} \label{eta_stirling}
\nonumber \eta_{Stirling} 
\nonumber &= -\frac{T_H \ln \left[ \frac{Z_2(E_{n,2})}{Z_1(E_{n,1})} \right] - T_C \ln \left[ \frac{Z_4(E_{n,4})}{Z_3(E_{n,3})}\right]}{ T_H^2 \frac{\partial}{\partial T_H} \ln\left[ Z_{2}(E_{n,2})\right] +  T_H \ln \left[ \frac{Z_2(E_{n,2})}{Z_1(E_{n,1})} \right] -  T_C^2 \frac{\partial }{\partial T_C}\ln\left[Z_{4}(T_C)\right]}.
\end{align}}
\end{widetext}

It is worth mentioning that many theoretical investigations on different quantum systems \cite{su2018carnot, thomas2019quantum, chatterjee2021temperature, purkait2022performance, das2023quantum, xia2024performance} operating under Stirling cycles reaches the classical Carnot efficiency given by \ref{eta_Carnot_clas} when imposed the low-temperature limit ($T_H \rightarrow T_C$).

\subsubsection{Quantum Otto Cycle} 

\begin{figure}[ht]
		\centering
		\includegraphics[width=0.5\textwidth]{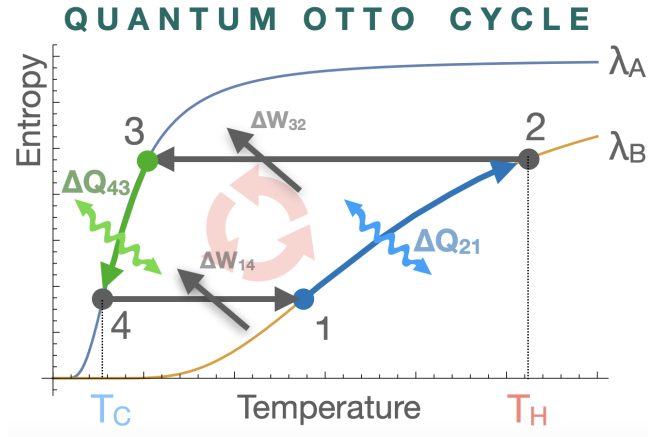}
		\caption{Entropy - Temperature S - T diagram of a quantum Otto cycle. Note that the work is performed by the system ($W>0$) during the adiabatic cooling ($2 \rightarrow 3$) process, while it is performed on the system ($W<0$) during the adiabatic heating ($3 \rightarrow 4$) process, similarly to the quantum Carnot process. The heat exchange between the working substance and the reservoir occurs only during the adiabatic}
    \label{otto}
\end{figure}

We define the quantum Otto cycle by combining two quantum adiabatic and two isochoric processes. Similar to the Carnot cycle, there is only heat exchange and work done for each stroke, opening the possibility of understanding the nature of both quantities independently. Figure \ref{otto} illustrates the quantum Otto cycle, emphasizing the heat fluxes by the working substance and the corresponding work done \textit{on} or \textit{by} the system.

The net work performed during the Otto cycle is given by adding up the work performed along the adiabatic processes displayed in Figure \ref{otto}:
\begin{align} 
  \Delta W_{Otto} =& \Delta W_{32} - \Delta W_{14} \\
   =&  k_B T_3^2 \frac{\partial }{\partial T_3}\ln\left[Z_{3}(T_3,E_{n,3})\right]- k_B T_h^2 \frac{\partial }{\partial T_h}\ln \left[Z_{2}(T_h,E_{n,2})\right] \\&- k_B T_1^2 \frac{\partial }{\partial T_1}\ln \left[Z_{1}(T_1,E_{n,1})\right] + k_B T_c^2 \frac{\partial }{\partial T_c}\ln\left[Z_{4}(T_c,E_{n,4})\right].
\end{align}
where $T_1$ and $T_3$ stand for the temperature at points $1$ and $3$, respectively. Again, we draw attention for the fact that we used the signal convention defined in Table \ref{tab:modes} for the work on each stroke.

\begin{widetext}
The efficiency of the Quantum Otto cycle can also be obtained from the above quantities:
\begin{eqnarray} \label{eta_otto}
     \nonumber \eta_{Otto} 
     &= \frac{ - \left\{ T_3^2 \frac{\partial }{\partial T_3}\ln\left[Z_{3}(T_3,E_{n,3})\right]-  T_H^2 \frac{\partial }{\partial T_H}\ln \left[Z_{2}(T_H,E_{n,2})\right] - T_1^2 \frac{\partial }{\partial T_1}\ln \left[Z_{1}(T_1,E_{n,1})\right] +  T_C^2 \frac{\partial }{\partial T_C}\ln\left[Z_{4}(T_C,E_{n,4})\right] \right\} }{ T_H^2 \frac{\partial }{\partial T_H}\ln\left[Z_{2}(T_H)\right]- T_1^2 \frac{\partial }{\partial T_1}\ln\left[Z_{1}(T_1)\right]} 
\end{eqnarray}
\end{widetext}

It is important to note that we do not recover the classical Carnot limit from the Quantum Otto cycle. In fact, for a two-level quantum system operating under the Otto cycle, the efficiency depends on the ratio between the level spacings during the two isochoric processes, while for a harmonic oscillator, it depends on the frequency $\frac{\omega_{H}}{\omega_{C}}$ between the hot and cold baths \cite{Quan:07}.

\section{Literature highlights} \label{sec:LitRes}

Modeling quantum heat engines from metal complex systems through quantum thermodynamics theory opens up a whole new horizon for exploring the technological potential of these materials. In this section, we discuss collected results from the literature that illustrate the versatility and relevance of this approach as a tool for establishing work extraction techniques on metal complex compounds.


\subsection{Magnetic field effects on quantum heat engines and refrigerators: toy models for metal complexes} \label{LitRes:MagQuanMach}

In addition to pressure, the applied magnetic field significantly influences the quantum thermodynamics of interacting spin systems, which are used to model metal complex compounds \cite{ccakmak2020construction, dyszel2021thermodynamics, ghannadan2021magnetic,azimi2022entanglement}, therefore playing a central role on Heinseberg spin based quantum thermal machines

Çakmak et al.  investigated a quantum Carnot cycle (Figure \ref{carnot}), employing various spin systems as working substances. In their study, the operation mode for the single spin and two uncoupled spins cases functions as a heat engine, with efficiency constrained by the classical Carnot limit, as given by equation \ref{eta_Carnot_clas}. However, introducing interactions between the pair of spin-$1/2$ particles via a coupling magnetic constant $J$ results in reduced efficiency and compromises the reversibility of the cycle. These findings are noteworthy from an experimental perspective, as they indicate that weakening intracluster interactions is a critical factor when considering dimeric metal complex compounds as working substances for realistic quantum Carnot cycles. Furthermore, Thomas et al. \cite{thomas2011coupled} and V. Mehta et al. \cite{mehta2017quantum} each identified an upper bound for the efficiency that is more restrictive than the Carnot limit when modeling the same dimeric system. Both studies concluded that this upper bound is determined solely by the control parameters of the Hamiltonian, independent of the temperatures of the hot and cold thermal baths.

The quantum Stirling cycle (Figure \ref{stirling}) has also been extensively used to explore the thermodynamics of spin configurations used to describe complex metal systems. For instance, Huang et al. \cite{huang2014quantum} examined single and interacting spin-$1/2$ systems under the influence of an external magnetic field $B$. Among their most intriguing results are the system's ability to surpass the Carnot efficiency for a single spin and the role of the coupling constant $J$ in switching the operation mode between a heat engine and a refrigerator (refer to Figure \ref{tab:modes} for a visual representation). More recently, Araya et al. \cite{araya2023magnetic} modeled quantum Stirling cycles for two interacting spin-$1/2$ particles under the influence of both a magnetic field $B$ and anisotropy $K$, demonstrating that the anisotropic system exhibits higher efficiency than its isotropic counterpart. A. Hassan et al. \cite{pili2023quantum} analyzed the same system using an XYZ Heisenberg model, incorporating a Dzyaloshinskii–Moriya interaction term in the Hamiltonian. In this case, the system exhibited two distinct operational modes—heat engine and refrigerator—depending on the appropriate selection of the magnetic field strength and the thermal bath temperature ratio.

These results underscore the influence of the applied magnetic field $B$, the temperatures of the thermal baths $T_H$ and $T_C$, and the intracluster interaction $J$ on the thermodynamic properties of spin systems operating under a quantum Stirling cycle.

Additionaly, recent works used the quantum thermodynamic approach to investigate the role of applied magnetic fields and anisotropy on quantum entanglement \cite{cruz2024entanglement, wang2024critical} and quantum phase transitions \cite{ghannadan2021magnetic, vargova2021unconventional} of Heisenberg-like spin systems. These approach are crucial for understading how to manipulate external parameters to harness quantum features for work extraction purposes.

\subsection{Quantum heat engines in metal complexes: caloric effects and thermodynamic implications}

\begin{figure}[h]
		\centering
		\includegraphics[scale=0.65]{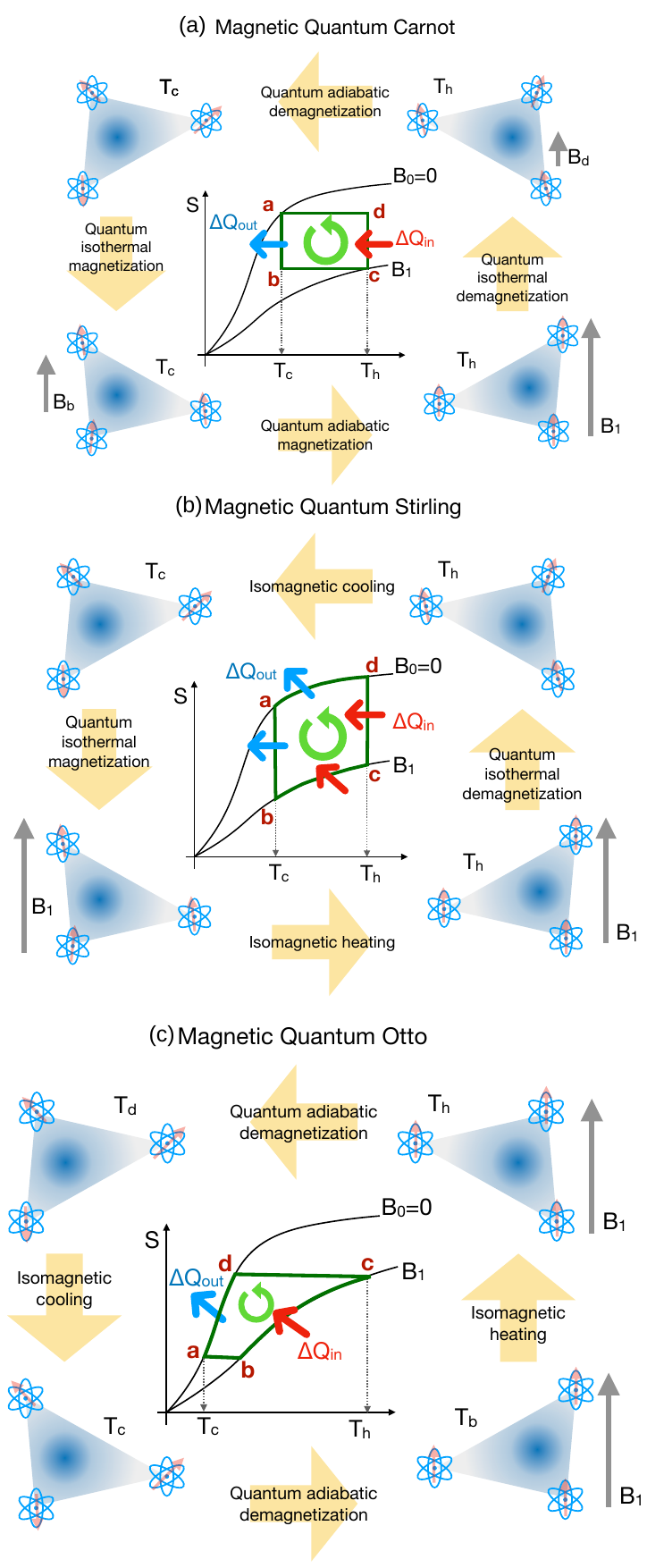}
		\caption{Schematic representation of a magnetic Quantum (a) Carnot, (b) Stirling, and (c) Otto machine. The action of the external perpendicular field $B_{1}$ on the triangular lattice of interacting spins is highlighted. Adapted from reference \cite{rojas2024quantum}.}
   \label{mag_quan_mach}
\end{figure}

Following the methodology outlined in section \ref{sec:QuanThermCyc}, O. Rojas et al. \cite{rojas2024quantum} modeled magnetic Carnot, Stirling, and Otto quantum cycles (Figures \ref{mag_quan_mach}(a), \ref{mag_quan_mach}(b), and \ref{mag_quan_mach}(c), respectively) using a trimeric spin configuration as the working substance, subject to perpendicular ($B_1$) and parallel ($B_0$) magnetic fields as external control parameters. It is important to note that, in this case, the external magnetic fields do not influence the magnetic coupling constants of the spins, unlike the scenarios discussed earlier in sections \ref{sec:QuanThermProc} and \ref{sec:QuanThermCyc}. This specific spin configuration can be used to describe the metal complex compounds $Cu_3$-$X$ (X = As, Sb), which, according to the literature \cite{binz2006theory, choi2008pulsed, shirata2012experimental}, exhibit an isosceles triangular spin ring and a cryogenic magnetocaloric effect (MCE) with an ordering temperature around 1.0 K. The authors primarily discuss the distinct operational modes that emerge as the values of $B_0$ and $B_1$ are tuned across both negative and positive regimes, demonstrating that both the Stirling and Otto cycles can operate under various conditions.

For the Carnot machine modeled, the system operates as a heat engine when $B_1 \leq B_0$ and as a refrigerator when $B_1 \geq B_0$, showing a notable change in operating mode near the critical temperature, where it functions as a heat engine even for $B_1 \geq B_0$. Similarly, the quantum Otto machine displays versatility in its modes of operation, predominantly acting as a thermal accelerator for $B_1 < B_0$, while primarily functioning as a refrigerator for $B_1 > B_0$. Interestingly, it can switch between the heat engine, refrigerator, heater, and thermal accelerator modes, depending on the specific combination of $T_H$, $T_C$, $B_0$, and $B_1$. A narrow region for $B_1 \geq B_0$ is also found, where it operates as a heat engine. The quantum Stirling machine shows similar flexibility, being capable of functioning as a heat engine, thermal accelerator, heater, or refrigerator. It behaves predominantly as a heat engine for $B_1 > B_0$ and as a refrigerator for $B_1 < B_0$. Near the MCE, the system alternates between a heat engine and a thermal accelerator, emphasizing the significance of the magnetocaloric effect. These findings indicate that modeling the quantum thermodynamic cycles of metal complexes must account for the influence of the cryogenic magnetocaloric effect on their quantum thermodynamic behavior.

The relationship between the magnetocaloric effect (MCE) and quantum properties relevant for modeling quantum cycles has been explored by various authors. In this context, Karlova et al. \cite{karǐova2017enhanced} studied noninteracting spin clusters using a spin-$1/2$ XXZ Heisenberg model, revealing unusual magnetic behavior attributed to quantum interactions, manifesting as magnetization plateaus accompanied by the MCE. Furthermore, T. Chakraborty et al. \cite{chakraborty2019magnetocaloric} investigated the link between MCE and quantum level crossing in spin-$1/2$ systems, collecting magnetization and specific heat data for the compound $NH_4CuPO_4 \cdot H_2O$, a spin-$1/2$ metal complex with antiferromagnetic coupling. By constructing entropy curves and calculating MCE values, the authors observed characteristics indicative of magnetic field-induced quantum level-crossing. Notably, recent work by Rojas et al. \cite{rojas2024quantum}, as discussed earlier in Section \ref{mag_quan_mach}, associates an anomaly in the predicted operational modes with the influence of the compound's MCE. These insights emphasize the critical role of the MCE in shaping the quantum thermodynamic behavior of spin systems.

\begin{figure}[ht]
		\centering
		\includegraphics[scale=0.5]{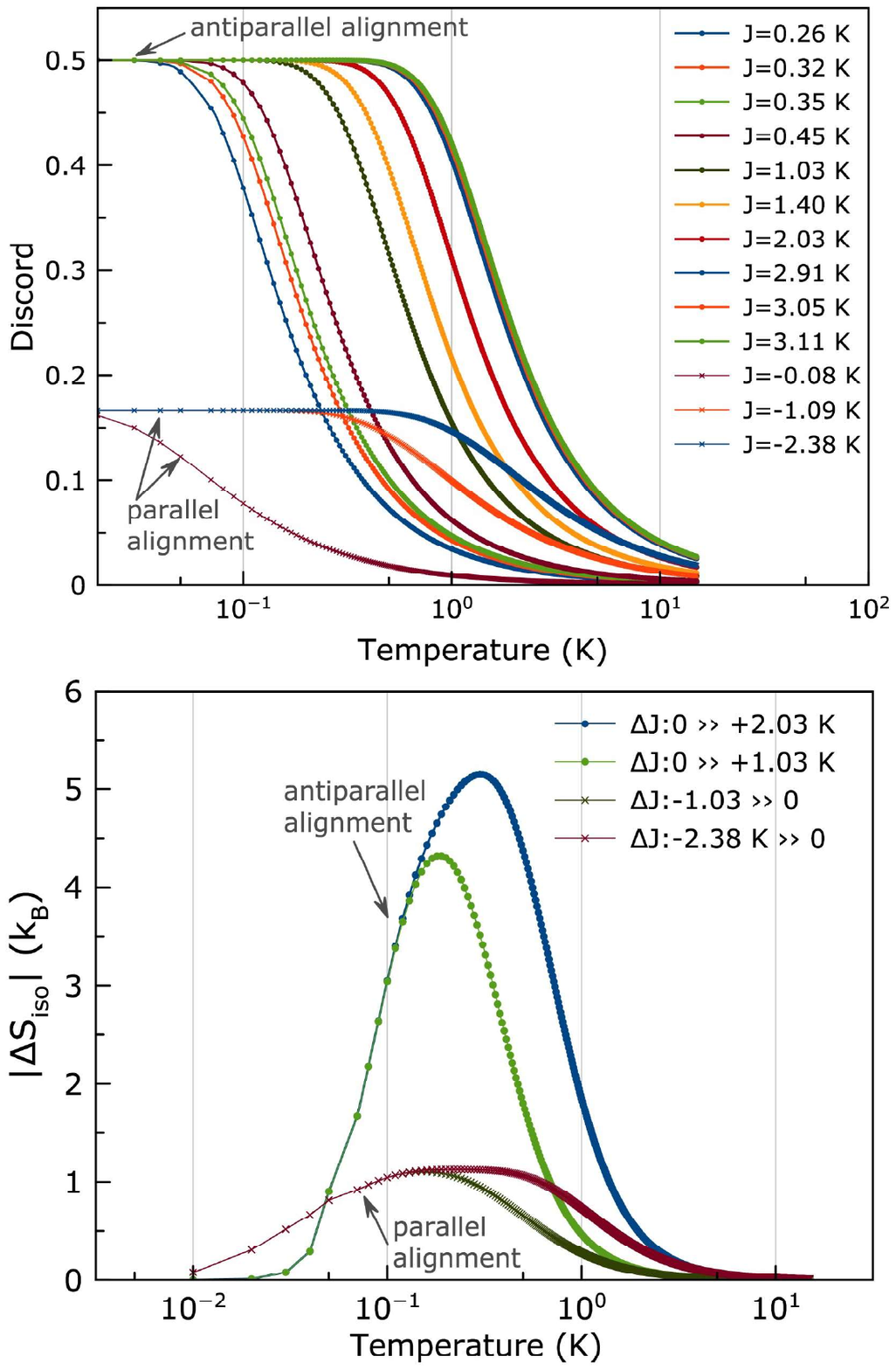}
		\caption{Correlation between magnetocaloric effect and quantum correlations. Here, both the thermal dependence of (a) the quantum discord $\mathcal{D}$ and (b) the corresponding $\Delta S_{iso}^{quan}$ for the $KNaCuSi_{4}O_{10}$, a dinuclear metal complex compound, were calculated after obtaining exchange parameters from first-principles methods as a function of pressure (up to 4.9 GPa). Reproduced from reference \cite{cruz2024quantum}with permission from the author.}
   \label{quantum_DS}
\end{figure}

In this context, a recently proposed approach was formulated by C. Cruz et al. \cite{cruz2024quantum}, introducing the concept of quantum caloric effects. In their work, the authors derive a general Maxwell relation based on a thermal average form of the Ehrenfest theorem, demonstrating that the isothermal entropy change can be intrinsically linked to the quantum correlations within the system. They specifically discuss the case of the $KNaCuSi_4O_{10}$ metal complex, where the variation in quantum discord $\mathcal{D}$ over the temperature range from $10^0$ K to $10^1$ K, as shown in Figure \ref{quantum_DS}(a), is associated with a peak in the entropy change, displayed in Figure \ref{quantum_DS}(b), within the same temperature interval. This observation highlights the profound connection between caloric effects and quantum correlations.

The observed variation in quantum discord $\mathcal{D}$ across the temperature range from $10^0$ K to $10^1$ K, illustrated in Figure \ref{quantum_DS}(a), corresponds to a peak in the isothermal entropy change $\Delta S_{iso}$ shown in Figure \ref{quantum_DS}(b). This peak arises due to changes in the occupation probabilities $p_n$ of the energy levels $E_n$, driven by variations in the system's magnetic coupling constant $J$. This approach is particularly relevant for the implementation of quantum heat engines using metal complex systems, as many materials in this class exhibit a magnetocaloric effect at cryogenic temperatures. Furthermore, given that metal complexes also exhibit barocaloric effects associated with spin-crossover transitions, it is a promising direction for future research to explore the role of quantum correlations in relation to the corresponding isothermal entropy change.


\newpage

\subsection{Experimental implementation of a quantum Stirling cycle and a quantum  batteries in metal complexes}\label{LitRes:StirMagPresData}

The experimental implementation of quantum thermodynamic cycles using metal complex systems as refrigerants is still in its early stages in the literature. Cruz et al. \cite{cruz2022} reported a recent advancement in this area. In their work, the authors constructed a quantum Stirling cycle, as depicted in Figure \ref{stirling}, for a dimeric system consisting of a pair of interacting spin-$1/2$ particles, closely following the mathematical framework described in section \ref{sec:QuanThermCyc}. Using hydrostatic pressure $P$ as an external parameter to tune the magnetic coupling constant $J$, both efficiency and work per cycle could be maximized. In addition, the system exhibited all four operational modes listed in Table \ref{tab:modes}: heating engine, accelerator, refrigerator, and heater.

Furthermore, the authors discovered that the quantum thermodynamic quantities of the system can be expressed as functions of the dimensionless magnetic susceptibility. This important finding establishes a connection between theoretical predictions and experimental investigations, which the authors in fact explored. Utilizing data from reference \cite{prescimone2010high}, they examined high-pressure experimental measurements of magnetic susceptibility as a function of applied pressure in a nuclear hydroxo-bridged Cu(II) compound with the formula $[Cu_{2}(OH){2}(bpy){2}][BF_4]_{2}$ (bpy = 2,2'-bipyridine). For the range of applied pressures used in the experiment, the Cu(II) compound was predicted to operate as a heat engine, regardless of the temperatures of the heat baths.

\begin{figure}[ht]
		\centering
		\includegraphics[scale=0.5]{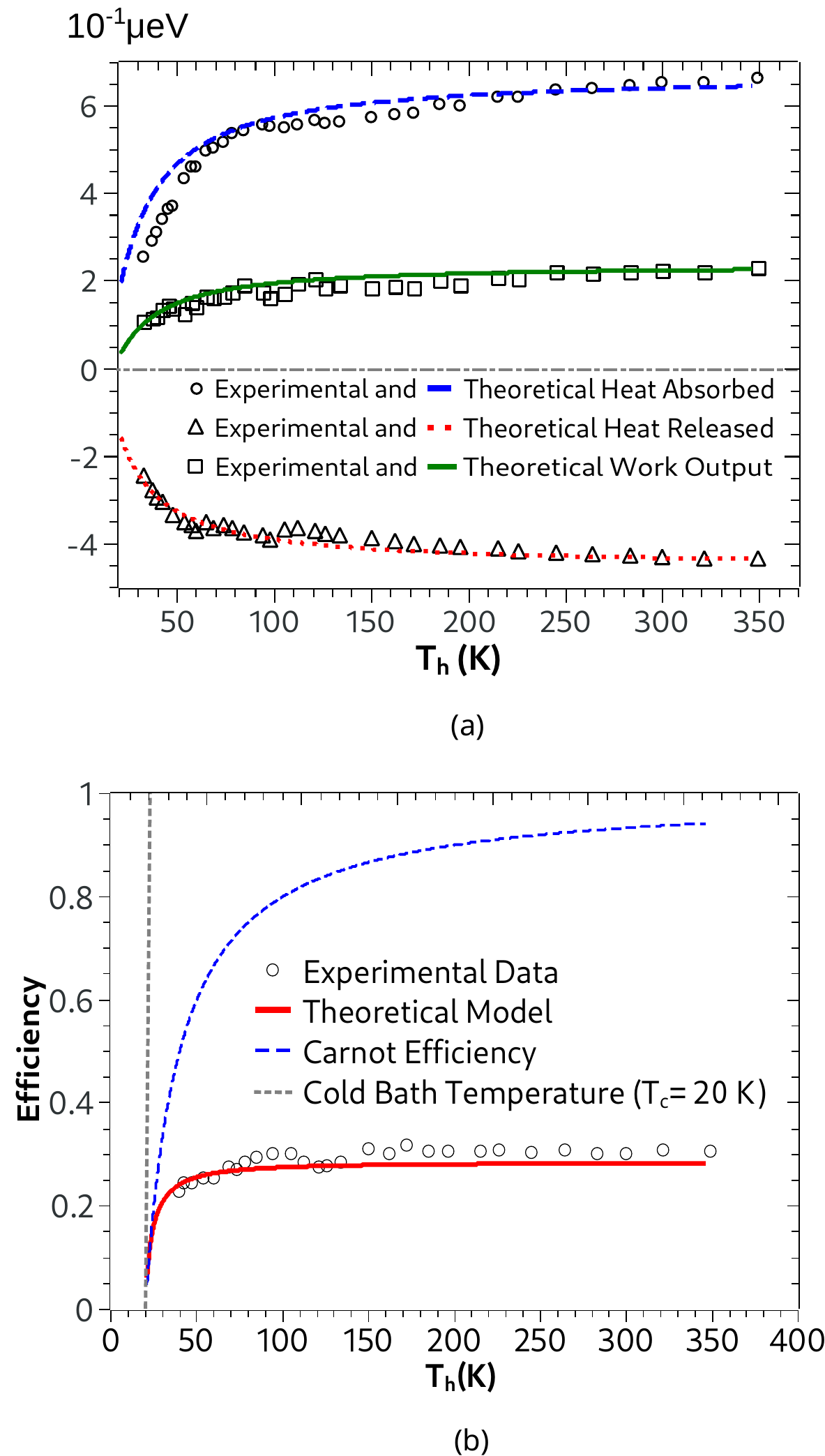}
		\caption{Comparison between theoretical thermodynamic quantities (a) work output, heat absorbed and released (heat fluxes) and (b) effciency (both as function of hot bath temperature $T_h$) calculated from theory for a dimeric system (empty points) and experimental ones (lines) obtained for a dinuclear hydroxo-bridged Cu(II) compound \cite{prescimone2010high}, considering a quantum Stirling machine. The engine operates between a hot reservoir at temperature $T_h$, and a cold one at fixed temperature $T_c = 20$ K (dotted gray line) - the lowest temperature measured for this system \cite{prescimone2010high}. The magnetic constant couplings are obtained from the Bleaney-Bowers fit of the magnetic
susceptibility, where $J_A/k_B = -42$ K is found for an applied pressure
of $P_A = 0.84$ GPa, and $J_B/k_B = -32$ K for the ambient pressure
($P_B = 10^{-4}$ GPa). Adapted from reference \cite{cruz2022}.}
   \label{exp_quan_stir}
\end{figure}

From Figure \ref{exp_quan_stir} (a), the experimental curves obtained from susceptibility data under pressure for the Cu(II) compound are well-fitted by the theoretical model that considers a dimeric spin-$1/2$ interacting system, demonstrating the suitability of this approach. Furthermore, Figure \ref{exp_quan_stir} (b) illustrates that the experimental efficiency of this Cu(II) metal complex approaches the classical Carnot limit when $T_h \rightarrow T_c $, as anticipated. The ability to express the quantum analogs of heat and work in terms of macroscopic classical quantities paves the way for the experimental implementation of quantum thermodynamic cycles using susceptibility data collected under varying pressures \cite{cruz2022}. 

On the other hand, in the study of quantum thermodynamic cycles, the concept of quantum batteries naturally emerges as a promising technological application for metal complex compounds. Quantum batteries are defined as energy storage devices that utilize quantum properties to store energy \cite{campaioli2023colloquium}, in contrast to conventional chemical batteries, which rely on chemical reactions to convert energy into electricity. The development and research of quantum batteries are actively progressing within the broader field of quantum technologies, aiming to enhance energy storage capabilities for diverse applications, ranging from quantum computing to renewable energy systems \cite{metzler2023emergence}.

\begin{figure}[ht]
		\centering
        \includegraphics[scale=0.5]{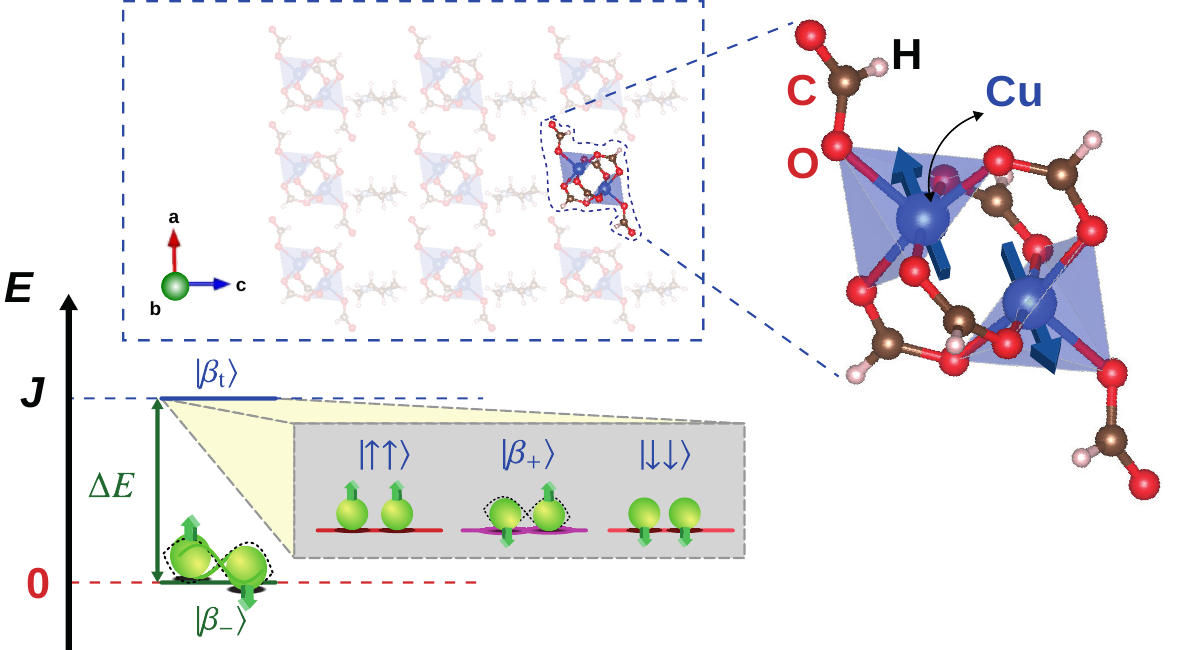}
		\caption{Sketch of the carboxylate-based metal complex Cu$_2$(HCOO)$_4$(HCOOH)$_2$(C$_4$H$_{10}$N$_2$), with a polyhedron representation. The compound has two internal energy levels: a singlet ground-state $\vert{\beta_{-}}\rangle$ and a triplet degenerate subspace ($\vert{\beta_{t}}\rangle$). The arrows in the sketch represent the effective spins that originate from the magnetic structure of the compound. Adapted from reference \cite{cruz2022}.}
		\label{quantum_battery}
	\end{figure}

Cruz et al. \cite{cruz2022quantum} explore the concept of a quantum battery that utilizes quantum discord as a resource for energy storage, focusing on a dinuclear metal complex with spin-$1/2$, specifically a metal-organic dinuclear copper(II) compound with the chemical formula Cu$_2$(HCOO)$_4$(HCOOH)$_2$(C$_4$H${_10}$N$_2$). The molecular structure of these carboxylate-based metal complexes provides resistance to external environmental disturbances such as temperature \cite{mario2,kova2020unconventional}, magnetic field \cite{kova2020unconventional,cruz2020quantifying,souza2}, and pressure \cite{cruz2017influence,cruz2020quantifying} effects. This shielding behavior is facilitated by the \textit{syn-syn} carboxylate bond between the metallic centers \cite{cruz2022quantum,souza,cruz2016carboxylate}, which creates a substantial energy gap ($\Delta E = J{\text{int}}/k{B} = 748$ K) between the ground state (a singlet entangled state) $\vert{\beta{-}}\rangle$ and the first excited states (the triplet separable subspace) ${\vert{\beta{\text{t}}}\rangle} = {\vert{\beta_{+}}\rangle, \vert{\downarrow\downarrow}\rangle, \vert{\uparrow\uparrow}\rangle}$, where $\vert{\beta_{\pm}}\rangle = (\vert{\downarrow\uparrow}\rangle \pm \vert{\uparrow\downarrow}\rangle)/\sqrt{2}$. Figure \ref{quantum_battery} illustrates the crystal structure of the compound, obtained through X-ray diffraction data \cite{cruz2016carboxylate}, along with the energy level configuration.

In this context, the quantum discord ($\mathcal{D}$) plays a crucial role, as the charge and discharge processes of the quantum battery can be directly associated with the corresponding increases and decreases in $\mathcal{D}$ values. Specifically, during the charging phase, an increase in quantum discord indicates that the system is gaining coherence and entanglement, thereby enhancing its capacity to store energy. Conversely, during the discharge phase, a decrease in quantum discord suggests a loss of these correlations, which is indicative of the energy being released.

The work extraction protocol is defined in terms of ergotropy $\mathcal{\epsilon}$, which represents the maximum amount of work that can be extracted from a quantum state via a unitary operation. This concept is pivotal in understanding the efficiency of quantum batteries, as it quantifies the extractable work based on the initial state of the system. These processes correspond to quantum isochoric processes, as elaborated in section \ref{sec:isochoric}, where the system undergoes energy exchanges without any changes in volume or temperature.

This analysis underscores the fundamental role of quantum correlations in shaping the quantum thermodynamic behavior of metal complex compounds. It emphasizes that quantum discord is not merely a theoretical construct, but a significant operational parameter that impacts the efficiency and effectiveness of quantum batteries. Furthermore, this study draws attention to ergotropy as a critical measure when considering such systems as quantum battery devices, providing insights into optimizing their performance for practical applications.

\section{Perspectives}\label{sec:Persp}

Quantum thermodynamics is a wide and fascinating field that offers profound insights into energy transfer and work at microscopic scales and is crucial for understanding how small-scale quantum systems behave and interact with their environments. Exploring small-scale quantum machines through a thermodynamic framework offers promising avenues for harnessing the quantum properties of real materials to enable work extraction. This approach helps to identify the role that genuine quantum resources play, such as quantum discord, in work extraction protocols, offering perspectives on how to leverage these properties to optimize the task.

For example, theoretical modeling of quantum magnetic cycles using Heisenber-like spin configurations as the working substance reveals the significant influence of magnetic fields and external pressure in modulating operational modes, net work output, and thermal efficiencies, through fine-tuning of the magnetic coupling constant $J$. These investigations reveal essential information regarding the enhancement of versatility and efficiency of quantum heat engines through the control of external stimuli.

It opens exciting new possibilities for exploiting the cryogenic magnetocaloric effect exhibited by metal complexes, which can enhance the performance and efficiency of quantum heat engines. Moreover, expanding this framework to include other caloric effects, such as barocaloric or electrocaloric effects, could significantly broaden the range of applications for metal complex compounds in quantum thermodynamic processes.

From an experimental point of view, the connection between magnetic susceptibility and quantum correlations recently discovered for dimeric metal complexes provides a practical pathway for implementing quantum cycles through straightforward measurements using commercial magnetometers. High-pressure susceptibility data from dimeric metal complexes, readily available in the literature, present an intriguing opportunity to advance experimental investigations in this field. This connection also suggests the potential for employing quantum thermodynamic processes to construct quantum batteries, where the work extraction can be quantified using ergotropy. Extending this approach toward more complex other spin configurations may be the key to uncovering new operational regimes and functionalities.

On the other hand, recent advancements in employing nuclear magnetic resonance (NMR) techniques to implement quantum Otto cycles have successfully demonstrated the operation of spin-based quantum heat engines \cite{pal2019experimental,lisboa2022experimental}, verified quantum fluctuation relations during cycles \cite{denzler2020efficiency,micadei2021experimental}, explored the role of quantum correlations such as quantum discord \cite{maziero2013quantum}, and even achieved the reversal of heat flow between systems \cite{micadei2019reversing}. A notable example is the work of J.P.S. et al. \cite{peterson2019experimental}, who utilized a $C^{13}$-labeled chloroform sample in an NMR spectrometer to perform a quantum Otto cycle, with $H^{1}$ nuclear spins mediating heat exchange. Although this experiment did not involve metal complex systems, the methodology could be adapted to electronic spin systems using electron paramagnetic resonance (EPR), a technique that enables precise determination of g-factors, crystal field parameters, and spin relaxation times in metal complexes \cite{moreno2021measuring, zadrozny2015millisecond, atzori2019second}. While NMR and EPR share foundational principles, such as the alignment of magnetic moments under an external field and the use of radiation to induce spin transitions, EPR's sensitivity to local magnetic environments due to the higher magnetic moment of electrons makes it particularly suitable for studying the intricate magnetic properties of metal complexes. Integrating EPR techniques for implementing quantum cycles could significantly expand the scope of experimental investigations, offering detailed insights into spin interactions and relaxation dynamics.

In summary, the critical assessment of the findings presented in this review highlights bold and innovative pathways for advancing the understanding of quantum behavior in metal complexes. These insights open up opportunities for developing efficient, sustainable, cost-effective, and widely accessible solutions for work extraction. By leveraging quantum resources, there is a real potential to revolutionize the energy sector and contribute to a more sustainable future through groundbreaking advancements in quantum thermodynamics.

\begin{acknowledgments}
VGP thanks Rio de Janeiro State Research Support Foundation (FAPERJ) for the financial support. The authors thank the Fundação de Amparo à Pesquisa do Estado da Bahia - FAPESB for its financial support (grant numbers APP0041/2023 and PPP0006/2024). This study was financed in part by the Coordenacao de Aperfeicoamento de Pessoal de Nivel Superior – Brasil (CAPES) – Finance Code 001. M.S.R. thanks FAPERJ and belongs to the INCT of Refrigeração e Termofísica, funding by CNPq by grant number 404023/2019-3. CICECO-Aveiro Institute of Materials, is supported by UIDB/50011/2020, UIDP/50011/2020 \& LA/P/0006/2020, financed by national funds through the FCT/MCTES (PIDDAC). 
\end{acknowledgments}

\newpage

\section{Appendix - Mathematical models} \label{sec:append}

\appendix

\section{Quantum Heat and Work} \label{AP:QuanHeatWork}

The internal energy of the system is the quantum-statistical average over the Hamiltonian:
\begin{equation} \label{AP:energy_U}
 U =Tr\{\hat{\rho}_{eq}\hat{H}\}=   \sum_{n} p_nE_n.
 \end{equation}
Differentiating equation \ref{energy_U}, we obtain:
\begin{equation} \label{AP:diff_energy_U}
  dU = \sum_{n} p_n dE_n + \sum_{n} E_n dp_n,
\end{equation}
which can be interpreted as the First Law of quantum thermodynamics. This can be compared with its classical counterpart:
\begin{equation} \label{AP:diff_First_Law}
  dU = \dbar W + \dbar Q  = \sum \mathcal{F}_id\mathcal{X}_i + TdS,
\end{equation}
where $S$ refers to the entropy, $\mathcal{X}_i$ is the generalized coordinate and $\mathcal{F}_i$ is the corresponding force. 

Following the definition provided by Alick \cite{Alicki:79} the first term of the equation \ref{diff_energy_U} is the quantum work done $dW$:
\begin{equation} \label{AP:quantum_dW}
  dW = \sum_{n} p_n dE_n.
\end{equation}
while changes the occupancy probability of each energy level $p_n$ correspond to a quantum heat exchange $dQ$, the second term of the equation \ref{diff_energy_U}:
\begin{equation} \label{AP:quantum_dQ}
  dQ = \sum_{n} E_n dp_n,
\end{equation}
where $dQ>0$ ($dQ<0$) means heat absorbed (released) by the system, adopting the point of view from the working substance.

\section{Quantum Carnot cycle} \label{AP:QuanCarn}

The amount of heat exchanged during isothermal strokes $1\rightarrow2$ and $3\rightarrow4$ of the quantum Carnot cycles displayed in Figure \ref{carnot} are given by equation \ref{Q_21_isothermal} and can be written as: 

\begin{equation} \label{AP:Q_21_carnot}
        \Delta Q_{21} =k_B T_H^2 \frac{\partial}{\partial T_H} \ln\left[ \frac{Z_{2}(E_{n,2})}{Z_{1}(E_{n,1})}\right] -k_B T_H \ln \left[ \frac{Z_2(E_{n,2})}{Z_1(E_{n,1})} \right],
\end{equation}
and
\begin{equation} \label{AP:Ap_Q_43_carnot}
    \Delta Q_{43} =k_B T_C^2 \frac{\partial}{\partial T_C} \ln\left[ \frac{Z_{4}(E_{n,4})}{Z_{3}(E_{n,3})}\right]  -k_B T_C \ln \left[ \frac{Z_4(E_{n,4})}{Z_3(E_{n,3})} \right] .
\end{equation}

On the other hand, the quantum adiabatic processes do not produce heat; therefore the work associated with these two strokes are described by equation \ref{W_21_adiabatic} and we write:
\begin{align} \label{AP:W_32_carnot}
    \Delta W_{32} &= U_3 - U_2 \\
    &= k_B T_C^2 \frac{\partial }{\partial T_C}\ln\left[Z_{3}(T_C,E_{n,3})\right]- k_B T_H^2 \frac{\partial }{\partial T_H}\ln \left[Z_{2}(T_H,E_{n,2})\right].
\end{align}

and 
\begin{align} \label{AP:W_14_Carnot}
    \Delta W_{14} &= U_1 - U_4 \\
    &= k_B T_H^2 \frac{\partial }{\partial T_H}\ln \left[Z_{1}(T_H,E_{n,1})\right]-k_B T_C^2 \frac{\partial }{\partial T_C}\ln\left[Z_{4}(T_C,E_{n,4})\right].
\end{align}

For the Carnot cycle, the network during the cycle is computed by summing up the contributions from each stroke process:

\begin{align} \label{AP:Wnet_Carnot}
\Delta W_{Carnot} 
&= \left(\Delta W_{21} - \Delta W_{43}\right)+ \left(\Delta W_{32} + \Delta W_{14}\right) \\
&= k_B T_H \ln \left[ \frac{Z_2(E_{n,2})}{Z_1(E_{n,1})} \right] -k_B T_C \ln \left[ \frac{Z_4(E_{n,4})}{Z_3(E_{n,3})} \right]\\
&+k_B T_C^2 \frac{\partial }{\partial T_C}\ln\left[Z_{3}(T_C,E_{n,3})\right]- k_B T_H^2 \frac{\partial }{\partial T_H}\ln \left[Z_{2}(T_H,E_{n,2})\right] \\
&+ k_B T_H^2 \frac{\partial }{\partial T_H}\ln \left[Z_{1}(T_H,E_{n,1})\right]-k_B T_C^2 \frac{\partial }{\partial T_C}\ln\left[Z_{4}(T_C,E_{n,4})\right]  \\ 
\end{align}

For the Carnot cycle, the quantities $\Delta Q_{21}$ and $\Delta Q_{43}$ displayed in Figure \ref{Q_21_isothermal} correspond to $\Delta Q_{in}$ and $\Delta Q_{out}$ heat fluxes, respectively. Then, we can also recover the network given by equation \ref{AP:Wnet_Carnot} by adding the heat fluxes $Q_{in}$ and $Q_{out}$:

\begin{align} \label{AP:1law_quantum_carnot}
    \Delta Q_{in} + \Delta Q_{out}&=\Delta Q_{21} + \Delta Q_{43} \\
    &=k_B T_H^2 \frac{\partial}{\partial T_H} \ln\left[ \frac{Z_{2}(E_{n,2})}{Z_{1}(E_{n,1})}\right] -k_B T_H \ln \left[ \frac{Z_2(E_{n,2})}{Z_1(E_{n,1})} \right]+ k_B T_C^2 \frac{\partial}{\partial T_C} \ln\left[ \frac{Z_{4}(E_{n,4})}{Z_{3}(E_{n,3})}\right] - k_B T_C \ln \left[ \frac{Z_4(E_{n,4})}{Z_3(E_{n,3})} \right]\\
    &=-\Delta W_{Carnot},
\end{align}

which verifies the validity of the 1º Law of Thermodynamics for the quantum Stirling cycle. 

As consequence of the adiabatic theorem given by equation \ref{constraint_carnot} it is easy to show that the Quantum Carnot efficiency is bounded by the classical Carnot limit:
\begin{align} \label{AP:eta_Carnot_clas}
    \eta_{Carnot}&= 1 + \left( \frac{T_C}{T_H}\right)\frac{T_C \overbrace{\frac{\partial}{\partial T_C} \ln\left[ \frac{Z_{4}(T_C,E_{n,4})}{Z_{3}(T_C,E_{n,3})}\right]}^{=0} + \ln \left[ \frac{Z_4(E_{n,4})}{Z_3(E_{n,3})} \right]}{T_H \underbrace{\frac{\partial}{\partial T_H} \ln\left[ \frac{Z_{2}(E_{n,2})}{Z_{1}(E_{n,1})}\right]}_{=0} - \ln \left[ \frac{Z_2(E_{n,2})}{Z_1(E_{n,1})} \right]} \\
   &=1 + \left( -\frac{T_C}{T_H}\right)\frac{\ln \left[ \frac{Z_4(E_{n,4})}{Z_3(E_{n,3})} \right]}{\ln \left[ \frac{Z_2(E_{n,2})}{Z_1(E_{n,1})} \right]} =1 - \frac{T_C}{T_H}
\end{align}

given by equation \ref{eta_Carnot_clas}.

\section{Quantum Stirling cycle}\label{AP:QuanStir}

During the isothermal process $1 \rightarrow 2$ and $3 \rightarrow 4$ of the quantum Stirling cycle exhibited by Figure \ref{stirling} the heat exchange is obtained using equation \ref{Q_21_isothermal_2} and written as:
\begin{equation} \label{AP:Q_21_stirling}
        \Delta Q_{21} =\underbrace{k_B T_H^2 \frac{\partial}{\partial T_H} \ln\left[ \frac{Z_{2}(E_{n,2})}{Z_{1}(E_{n,1})}\right]}_{\Delta U_{21}}- \underbrace{k_B T_H \ln \left[ \frac{Z_2(E_{n,2})}{Z_1(E_{n,1})} \right]}_{\Delta W_{21}},
\end{equation}
and
\begin{equation} \label{AP:Q_43_stirling}
    \Delta Q_{43} =\underbrace{k_B T_C^2 \frac{\partial}{\partial T_C} \ln\left[ \frac{Z_{4}(E_{n,4})}{Z_{3}(E_{n,3})}\right]}_{\Delta U_{43}}  -\underbrace{k_B T_C \ln \left[ \frac{Z_4(E_{n,4})}{Z_3(E_{n,3})} \right]}_{\Delta W_{43}} .
\end{equation}
which are identical of the ones obtained for the Carnot case, since they are describing the same quantum isothermal process.

On the other hand, for the two isochoric processes $2 \rightarrow 3$ and $4 \rightarrow 1$ no work is done, and the heat exchanged during these two processes can be obtained using equation \ref{Q_21_isochoric}, explicitaly written as: 
\begin{equation} \label{AP:Q_32_stirling}
    \Delta Q_{32} = k_B T_C^2 \frac{\partial }{\partial T_C}\ln\left[Z_{3}(T_C)\right]- k_B T_H^2 \frac{\partial }{\partial T_H}\ln\left[Z_{2}(T_H)\right],
\end{equation}
and 
\begin{equation} \label{AP:Q_14_stirling}
    \Delta Q_{14} = k_B T_H^2 \frac{\partial }{\partial T_H}\ln\left[Z_{1}(T_H)\right]- k_B T_C^2 \frac{\partial }{\partial T_C}\ln\left[Z_{4}(T_C)\right] .
\end{equation}

For the Stirling cycle, the heat fluxes are defined as: 

\begin{align} \label{AP:Q_in_Stirling}
\Delta Q_{in}=\Delta Q_{21} + \Delta Q_{14} 
&=k_B T_H^2 \frac{\partial}{\partial T_H} \ln\left[ \frac{Z_{2}(E_{n,2})}{Z_{1}(E_{n,1})}\right] -k_B T_H \ln \left[ \frac{Z_2(E_{n,2})}{Z_1(E_{n,1})} \right] \\
&+ k_B T_H^2 \frac{\partial }{\partial T_H}\ln\left[Z_{1}(T_H)\right]- k_B T_C^2 \frac{\partial }{\partial T_C}\ln\left[Z_{4}(T_C)\right] ,
\end{align}
and
\begin{align} \label{AP:Q_out_Stirling}
\Delta Q_{out}= \Delta Q_{43} + \Delta Q_{32}
&= k_B T_C^2 \frac{\partial}{\partial T_C} \ln\left[ \frac{Z_{4}(E_{n,4})}{Z_{3}(E_{n,3})}\right]  -k_B T_C \ln \left[ \frac{Z_4(E_{n,4})}{Z_3(E_{n,3})} \right]\\
&+ k_B T_C^2 \frac{\partial }{\partial T_C}\ln\left[Z_{3}(T_C)\right]- k_B T_H^2 \frac{\partial }{\partial T_H}\ln\left[Z_{2}(T_H)\right] .
\end{align}

The net Work due is easily computed by summing up to the contribution from the two isothermal processes:
\begin{align} \label{AP:W_net_Stirling}
\Delta W_{Stirling} &= \Delta W_{21} -\Delta W_{43}\\
&= k_B T_H \ln \left[ \frac{Z_2(E_{n,2})}{Z_1(E_{n,1})} \right] -k_B T_C \ln \left[ \frac{Z_4(E_{n,4})}{Z_3(E_{n,3})} \right]
\end{align}

where we used the convention defined in Table \ref{tab:modes}: $\Delta W_{43}<0$ since it accounts for the Work performed on the system (see Figure \ref{stirling} for clarity).

For completeness, net Work for one entire cycle obtained by adding $Q_{in}$ and $Q_{out}$ recovers the one as the sum of the works performed on/by the system:

\begin{align} \label{AP:1law_quantum_stirling}
    \Delta Q_{in} + \Delta Q_{out} 
&=k_B T_H^2 \frac{\partial}{\partial T_H} \ln\left[ \frac{Z_{2}(E_{n,2})}{Z_{1}(E_{n,1})}\right] -k_B T_H \ln \left[ \frac{Z_2(E_{n,2})}{Z_1(E_{n,1})} \right] \\
&+ k_B T_H^2 \frac{\partial }{\partial T_H}\ln\left[Z_{1}(T_H)\right]- k_B T_C^2 \frac{\partial }{\partial T_C}\ln\left[Z_{4}(T_C)\right] \\
&+ k_B T_C^2 \frac{\partial}{\partial T_C} \ln\left[ \frac{Z_{4}(E_{n,4})}{Z_{3}(E_{n,3})}\right]  -k_B T_C \ln \left[ \frac{Z_4(E_{n,4})}{Z_3(E_{n,3})} \right]\\
&+ k_B T_C^2 \frac{\partial }{\partial T_C}\ln\left[Z_{3}(T_C)\right]- k_B T_H^2 \frac{\partial }{\partial T_H}\ln\left[Z_{2}(T_H)\right] \\
&= -\left( k_B T_H \ln \left[ \frac{Z_2(E_{n,2})}{Z_1(E_{n,1})} \right] -k_B T_C \ln \left[ \frac{Z_4(E_{n,4})}{Z_3(E_{n,3})} \right] \right) =-\Delta W_{Stirling}
\end{align}

as expected by the 1º Law of Thermodynamics.

The efficiency of the Stirling cycle is given by equation \ref{eta_def} and is written as:
\begin{widetext}
{\begin{align} \label{AP:Eff_Stirling}
\nonumber \eta_{Stirling} &=-\frac{\Delta W_{Stirling}}{\Delta Q_{in}} \\
&=- \frac{k_B T_H \ln \left[ \frac{Z_2(E_{n,2})}{Z_1(E_{n,1})} \right] -k_B T_C \ln \left[ \frac{Z_4(E_{n,4})}{Z_3(E_{n,3})} \right]}{k_B T_H^2 \frac{\partial}{\partial T_H} \ln\left[ \frac{Z_{2}(E_{n,2})}{Z_{1}(E_{n,1})}\right] -k_B T_H \ln \left[ \frac{Z_2(E_{n,2})}{Z_1(E_{n,1})} \right] + k_B T_H^2 \frac{\partial }{\partial T_H}\ln\left[Z_{1}(T_H)\right]- k_B T_C^2 \frac{\partial }{\partial T_C}\ln\left[Z_{4}(T_C)\right]} \\
&= \frac{T_H \ln \left[ \frac{Z_2(E_{n,2})}{Z_1(E_{n,1})} \right] - T_C \ln \left[ \frac{Z_4(E_{n,4})}{Z_3(E_{n,3})}\right]}{ T_H^2 \frac{\partial}{\partial T_H} \ln\left[ Z_{2}(E_{n,2})\right] +  T_H \ln \left[ \frac{Z_2(E_{n,2})}{Z_1(E_{n,1})} \right] -  T_C^2 \frac{\partial }{\partial T_C}\ln\left[Z_{4}(T_C)\right]}.
\end{align}}
\end{widetext}

\section{Quantum Otto cycle} \label{AP:QuanOtto}

For the Otto cycle the strokes $2\rightarrow3$ and $4\rightarrow1$, corresponding to quantum adiabatic processes do not produce heat; the corresponding work for these processes are written accordingly to equation \ref{W_21_adiabatic}:
\begin{align} \label{AP:W_32_otto}
    \Delta W_{32} &=\Delta U_{32}=U_{3} - U_{2} \\
    &= k_B T_3^2 \frac{\partial }{\partial T_3}\ln\left[Z_{3}(T_3,E_{n,3})\right]- k_B T_H^2 \frac{\partial }{\partial T_H}\ln \left[Z_{2}(T_H,E_{n,2})\right].
\end{align}
and 
\begin{align} \label{AP:W_14_otto}
    \Delta W_{14} &= \Delta U_{14} = U_{1} - U_{4} \\
    &= k_B T_1^2 \frac{\partial }{\partial T_1}\ln \left[Z_{1}(T_1,E_{n,1})\right]-k_B T_C^2 \frac{\partial }{\partial T_C}\ln\left[Z_{4}(T_C,E_{n,4})\right]
\end{align}

In an analogous fashion, for the isochoric processes $1\rightarrow2$ and $3\rightarrow4$ the heats exchanged are given in terms of equation \ref{Q_21_isochoric}, and explicitly written as:
\begin{align} \label{AP:Q_21_otto}
    \Delta Q_{21} &= \Delta U_{21}=U_{2} - U_{1} \\
    &= k_B T_H^2 \frac{\partial }{\partial T_H}\ln\left[Z_{2}(T_H)\right]- k_B T_1^2 \frac{\partial }{\partial T_1}\ln\left[Z_{1}(T_1)\right],
\end{align}
and 
\begin{align} \label{AP:Q_43_otto}
    \Delta Q_{43} &= \Delta U_{43}=U_{4} - U_{3} \\
    &= k_B T_c^2 \frac{\partial }{\partial T_c}\ln\left[Z_{4}(T_c)\right]- k_B T_3^2 \frac{\partial }{\partial T_3}\ln\left[Z_{3}(T_3)\right] .
\end{align}
where $T_1$ and $T_3$ stand for the temperature at points $1$ and $3$, respectively.

The net work performed during the Otto cycle is given by adding up the work performed along the adiabatic processes:
\begin{align} \label{AP:net_work_otto}
  \Delta W_{Otto} &= \Delta W_{32} - \Delta W_{14} \\
   &=  k_B T_3^2 \frac{\partial }{\partial T_3}\ln\left[Z_{3}(T_3,E_{n,3})\right]- k_B T_h^2 \frac{\partial }{\partial T_h}\ln \left[Z_{2}(T_h,E_{n,2})\right] \\&- k_B T_1^2 \frac{\partial }{\partial T_1}\ln \left[Z_{1}(T_1,E_{n,1})\right] + k_B T_c^2 \frac{\partial }{\partial T_c}\ln\left[Z_{4}(T_c,E_{n,4})\right] .
\end{align}

For the Otto cycle, the heat fluxes $Q_{in}$ and $Q_{out}$ can be identified to the $\Delta Q_{21}$ and $\Delta Q_{43}$ quantities above calculated. Then, the net Work can be obtained by :

\begin{align} \label{AP:1law_quantum_otto}
    Q_{in} + Q_{out} &=\Delta Q_{21} + \Delta Q_{43} \\
    &= k_B T_H^2 \frac{\partial }{\partial T_H}\ln\left[Z_{2}(T_H)\right]- k_B T_1^2 \frac{\partial }{\partial T_1}\ln\left[Z_{1}(T_1)\right] \\
    &+ k_B T_c^2 \frac{\partial }{\partial T_c}\ln\left[Z_{4}(T_c)\right]- k_B T_3^2 \frac{\partial }{\partial T_3}\ln\left[Z_{3}(T_3)\right] \\
    &= - \Delta W_{Otto}
\end{align}

which verifies the 1º Law of thermodynamics for the Otto cycle.

Finally, the efficiency of the Quantum Otto cycle is obtained according to the definition given by equation \ref{eta_def}, and is written as:
\begin{align} \label{AP:eta_otto}
     \nonumber \eta_{Otto} &= \frac{-\Delta W_{Otto}}{Q_{in}} \\
     &= \frac{ - \left\{ T_3^2 \frac{\partial }{\partial T_3}\ln\left[Z_{3}(T_3,E_{n,3})\right]-  T_H^2 \frac{\partial }{\partial T_H}\ln \left[Z_{2}(T_H,E_{n,2})\right] - T_1^2 \frac{\partial }{\partial T_1}\ln \left[Z_{1}(T_1,E_{n,1})\right] +  T_C^2 \frac{\partial }{\partial T_C}\ln\left[Z_{4}(T_C,E_{n,4})\right] \right\} }{ T_H^2 \frac{\partial }{\partial T_H}\ln\left[Z_{2}(T_H)\right]- T_1^2 \frac{\partial }{\partial T_1}\ln\left[Z_{1}(T_1)\right]} \\
     &=1+\frac{  T_C^2 \frac{\partial }{\partial T_C}\ln\left[Z_{4}(T_C,E_{n,4})\right]  -  T_3^2 \frac{\partial }{\partial T_3}\ln\left[Z_{3}(T_3,E_{n,3})\right] }{ T_H^2 \frac{\partial }{\partial T_H}\ln\left[Z_{2}(T_H)\right]- T_1^2 \frac{\partial }{\partial T_1}\ln\left[Z_{1}(T_1)\right]}.
\end{align}

\end{document}